\documentclass[a4paper, 12pt]{article}
\pdfoutput=1

  \usepackage{a4wide}
  \usepackage{latexsym}
  \usepackage{epsf}
  \usepackage{amssymb}
  \usepackage{amsmath, cite}
  \usepackage{amsmath,amssymb,amsthm}
  \usepackage{verbatim}
  \usepackage{hypernat}

\usepackage[usenames,dvipsnames]{color}
\definecolor{MyDarkBlue}{rgb}{0,0.08,0.45}
\definecolor{LightBlue}{rgb}{0.684,0.999,0.999}
\usepackage{hyperref}
\hypersetup{
colorlinks=true,
citecolor=MyDarkBlue,
linkcolor=MyDarkBlue,
urlcolor=MyDarkBlue,
pdfauthor={Stefanos Katmadas and Ruben Minasian},
pdftitle={N=2 higher-derivative couplings from strings},
pdfsubject={hep-th}
}

\newcommand{\ft}[2]{{\textstyle\frac{#1}{#2}}}
\usepackage[titles]{tocloft}

\usepackage{slashbox}

\usepackage{rotating}

\def\Slash#1{\rlap{\hbox{$\mskip 3 mu /$}}#1}      
\numberwithin{equation}{section}

\newcommand{\nv}{n_\mathrm{v}}
\newcommand{\nn}{\nonumber}
\newcommand{\R}{\mbox{Re}}
\newcommand{\I}{\mbox{Im}}
\def\cH{{\mathcal H}}
\def\cR{{\mathcal R}}
\def\cK{{\mathcal K}}
\def\cV{{\mathcal V}}
\def\cN{{\mathcal N}}
\def\a{\alpha}
\def\b{\beta}

\newcommand\tr{\mathrm{tr\,}}

\newcommand{\HH}{\mathcal {H}}

\newcommand{\omp}{\Omega_+}
\newcommand{\omm}{\Omega_-}

\begin{document}
\pagenumbering{roman}
\begin{titlepage}
\titlepage

\begin{flushright}
IPhT-t13/250
\end{flushright}
\vspace{2.cm}

\begin{center}

\begin{center}
{\LARGE \bf{ $\cN\!=\!2$ higher-derivative couplings \\ from strings
}}
\end{center}

\vspace{1.1 cm} {\large Stefanos Katmadas$^{a}$ and Ruben Minasian$^b$} \\

\vspace{0.8 cm}{$^{a}$ Dipartimento di Fisica, Universit\'a di Milano-Bicocca, I-20126 Milano, Italy}\\
\vspace{.1 cm} {$^b$ Institut de Physique Th\'eorique, CEA
Saclay, \\CNRS URA 2306 ,  F-91191 Gif-sur-Yvette, France}
\\

\vspace{0.8cm}

{\bf Abstract}
\end{center}

\begin{quotation}

We consider the Calabi-Yau reduction of the Type IIA  eight derivative one-loop stringy corrections
focusing on the couplings of the four dimensional gravity multiplet with vector multiplets and
a tensor multiplet containing the NS two-form. We obtain a variety of higher derivative invariants
generalising the one-loop topological string coupling, $F_1$, controlled by the lowest order K\"ahler
potential and two new non-topological quantities built out of the Calabi-Yau Riemann curvature.

\end{quotation}

\vspace{7cm}
\flushbottom{{\ttfamily{\scriptsize $^{a}$stefanos.katmadas @ unimib.it, $^{b}$ruben.minasian @ cea.fr}}}
\end{titlepage}
\pagenumbering{arabic}

\tableofcontents

\section{Introduction and summary}

The quantum corrections in $\cN\!=\!2$ theories have received a great deal of attention.  These are of two types: corrections proportional to the inverse tension of the string and corrections proportional to the string coupling constant. The former arise from perturbative and instantonic world-sheet corrections and are encoded in higher derivative terms in the ten-dimensional supergravity action, while the latter come from string loops and brane instantons.  Perturbative low energy effective actions are expanded in a double perturbation series in the inverse tension and the coupling constant. 

These corrections not only affect the moduli spaces of $\cN\!=\!2$ theories, but  are manifested in the higher-derivative couplings.  A better control of these couplings is hence essential, as is demonstrated by the study of terms involving the Weyl chiral (supegravity) super field $W$.   However most of the other (higher-derivative) couplings in $\cN\!=\!2$ theories and their relation to string theory remain largely unexplored. We make some steps in this directions. Our study is mostly restricted to string one loop results and Calabi-Yau compactifications, and will not cover gauged $\cN\!=\!2$ theories.

\medskip
\noindent
The better-understood structures, involving $W$ holomorphically, are captured by the topological string theory. In particular, the F-term in the low energy effective action in four dimensions is related to the scattering amplitude of 2 selfdual gravitons and $(2g - 2)$ self-dual graviphotons in the zero-momentum limit and is computed by the genus-$g$ contribution $F_g$ to the topological string partition function  \cite{Bershadsky:1993cx, Antoniadis:1993ze}.  Crucially, the genus-$g$ contribution $F_g$ also determines the partition functions of $\cN\!=\!2$ global gauge theories. 

There exists, however, a continuous deformation of the gauge theory which uses nontrivially the manifest SU(2) R-symmetry of theory. This is what happens in the  so-called Omega background.  \cite{Nekrasov:2002qd, Nekrasov:2003rj}.The two-parameter gauge theory partition function in the Omega background has been computed recently and reduces to the standard gauge theory partition function only when the two parameters are set equal. It is an outstanding open problem to find string theory realisation of these backgrounds and understand the extension of the genus-$g$ function $F_g$  which determines the general $\cN\!=\!2$ partition function, and which should involve scattering amplitude among 2 gravitons, $(2g - 2)$ graviphotons, and $2n$ gauge fields in vector multiplets.  Theses considerations have lead to a recent interest in explicit realisation of couplings $F_{g,n} W^{2g} V^{2n}$ \cite{Morales:1996bp, Antoniadis:2010iq, Nakayama:2011be, Antoniadis:2013bja, Antoniadis:2013mna}.

Let us recall that the genus one partition function $F_1$ is special due to the fact that it is the only perturbative four-dimensional contribution, which survives the five-dimensional decompactification limit. The ten/eleven dimensional origin of these couplings is related to M5 brane anomalies and they lift to certain eight-derivative terms in the effective action \cite{Ferrara:1996hh, Antoniadis:1997eg}. Until very recently only the gravitational part of these couplings was known (and it was checked that their reduction on CY manifolds does correctly reproduce $F_1$). At present, we have a much better control of the more general form of the couplings in general string backgrounds with fluxes turned on, so that an explicit calculation of the one-loop four-, six- and eight-derivative couplings in $\cN\!=\!2$ theories, which should lead to the generalisation of $F_1$, is now within the reach.

\medskip
\noindent
In the four dimensional setting, recent developments in going beyond chiral couplings described by integrals over half of superpace \cite{deWit:2010za}, allow us to extend the list of higher derivative terms in several ways. The new couplings are constrained by $\cN\!=\!2$ supersymmetry to be governed by real functions of the four dimensional chiral fields. The latter naturally include vector multiplets and two types of chiral backgrounds, one of which is the Weyl background, $W^{2}$, introduced above. The second chiral background we consider is constructed out of the components of a tensor multiplet containing the NS two-form, the so called universal tensor multiplet\footnote{While there is no obstacle in considering a background of an arbitrary number of tensor multiplets in principle, we restrict our considerations to the universal tensor multiplet. We therefore ignore here all the complex deformations of the internal Calabi-Yau; including these in the reduction should yield couplings for generic 
hyper-
matter.} and contains four derivative terms on its 
components, such as $(\nabla H)^2$, where $H = d B$ and $B$ is the NS two-form. These ingredients then
allow us to describe couplings which are characterised by polynomials of the type $[F^2 + R^2 + (\nabla H)^2]^{n}$, generalising the purely gravitational $R^2$ couplings discussed above. The function of the vector multiplet scalars and Weyl background controlling these couplings directly corresponds to the extended couplings $F_{g,n} W^{2g} V^{2n}$, when the tensor multiplet is ignored. Inclusion of the latter results to more general couplings that have not yet been discussed in the literature.

\medskip
\noindent
From a quantum gravity point of view, higher-derivative corrections serve as a means of probing string theory at a fundamental level. Even though the complete expansion involves all fields of the theory, so far the attention has been mostly concentrated on the gravitational action. In particular, the one-loop eight derivative $R^4$ ($\mathcal{O}(\alpha'^3)$) terms stand out among the stringy quantum corrections.   Due to being connected to anomaly cancellation, they are not renormalised at higher loops  and survive the eleven-dimensional  strong coupling limit.  These couplings also play a special role in Calabi-Yau reductions to four-dimensional  $\cN\!=\!2$ theories. Firstly, they have been instrumental in understanding the perturbative corrections to the metrics on moduli spaces. 
In addition, they give rise to the four-derivative $R^2$ couplings, and as mentioned above agree with $F_1 W^2$.

In order to understand the stringy origin of more general higher derivative couplings in $\cN\!=\!2$ theories,
one needs to go beyond the purely gravitational couplings in ten dimensions.   In the NS-NS sector of string theory,  $H^2R^3$ couplings are  specified by a five-point function \cite{Peeters:2001ub}.  Direct amplitude calculations beyond this order are exceedingly difficult,  but recent progress in classification of string backgrounds using the generalised complex geometry and T-duality covariance provide rather powerful constrains on the structure of the quantum corrections in the effective actions.  A partial result for the six-point function, obtained recently, together with T-duality constraints and the heterotic/type II duality beyond leading order, allows to recover the ten-dimensional perturbative action almost entirely  \cite{Liu:2013dna} (the few yet unfixed terms mostly vanish in CY backgrounds and hence are not relevant for the current project). The eleven-dimensional lift of the modified coupling leads to the inclusion of the M-theory four-form field strength; the subsequent reduction on a non-
trivial KK monopole background allows to incorporate the full set of RR fields in the one-loop eight-derivative couplings. This knowledge will be crucially used for obtaining the relevant four-dimensional $\cN\!=\!2$ couplings.

\medskip
\noindent
The goal of this paper is to bring together some of these recent developments. In the process, we shall:
\begin{itemize}
\item[ ] Confirm and specify some of the predictions of general $\cN\!=\!2$ considerations and fix the a priori arbitrary quantities constrained solely by supersymmetry in terms of Calabi-Yau data
\item[ ] Discover new terms and couplings that have not been previously considered
\item[ ] Provide some tests and justification for the proposed lift of type IIA $R^4$ terms to eleven dimensions
\end{itemize}
A brief comment on the last point. Since the lift from ten to eleven dimensions involves a strong coupling limit, ones is normally suspicious of simple-minded arguments associated with just replacing the string theory NS three-form $H$ by  a four-form $G$. In $\cN\!=\!2$ theories however the three- and four-form give rise to fields in the same super multiplet, namely the (real part of the) scalars $u^I$ and the vectors $A^I$ in the vector matter respectively (here the index $I$ spans the vector multiplets). Hence verifying that the respective couplings involving $u^I$ ad $A^I$ are supersymmetric completions of each other provided a test of the lifting procedure.

\medskip
\noindent
We conclude this section by a summary of our results. A variety of four dimensional higher derivative
terms of the type $[F^2 + R^2 + (\nabla H)^2]^{n}$ are characterised by giving the relevant functions of
four dimensional chiral superfields that control them. From the point of view of the CY reduction, the
order of derivatives of all terms in four dimensions is controlled solely by the power of the CY Riemann
tensor appearing in the internal integrals. We therefore find that the eight, six and four derivative
terms are controlled by the possible integrals involving none, one or two powers of the internal Riemann tensor, respectively.

We find, in particular, that only the lowest order K\"ahler potential is relevant for the eight derivative
terms, since this is the natural real function of vector multiplet moduli arising in Calabi-Yau compactifications,
describing the total volume of the internal manifold.
At lower orders in derivatives, the K\"ahler potential still appears as part of the functions describing the various
invariants, combined with the Riemann tensor on the CY manifold $X$, denoted by $R_{mnpq}$. Given that all traces
of the latter vanish, the relevant internal integrals must necessarily contain the harmonic forms on the CY manifold.
In the case at hand, the relevant forms are the $h^{(1,1)}$ two-forms $\omega_I \in H^2(X, \mathbb{Z})$, where
$I,\, J = 1, \dots, h^{(1,1)}$, since we ignore the hypermultiplets arising from the $(2,1)$ cohomology.
We then obtain the following tensorial objects
\begin{gather}
\cR_{IJ} = \int_{X} R_{mnpq}\,\omega_{I}{}^{mn}\omega_{J}{}^{pq} \,,
\nn\\
X_{IJ} = \int_X \epsilon_{m n m_1 \ldots m_4} \epsilon_{pq n_1 \ldots n_4}
R^{m_1m_2n_1n_2}R^{m_3 m_4 n_3n_4} \,\,  \omega_I\,^{mn} \, \omega_J\,^{pq} \label{eq:R-expr}\,,
\end{gather}
which control all the couplings that we were able to describe within $\cN\!=\!2$ supergravity  at six- and four-derivative order respectively. Similar to the standard derivation of the lowest order K\"ahler potential, one can deduce the existence of corresponding real functions whose derivatives lead to the the couplings \eqref{eq:R-expr}. Finally, the inclusion of the Weyl and tensor superfields
through additional multiplicative factors leads to the functions that characterise the corresponding
couplings involving $R^2$ and $(\nabla H)^2$ respectively. For example, the $R^2 F^4$ and $R^2 F^2$
couplings lead to the functions
\begin{align}
R^2 (\nabla F)^2 \quad \Rightarrow \quad &\, A_{\sf w} \, \mathrm{e}^{2\,\cK(Y ,\bar{Y})}\,,
\nonumber\\
R^2 F^2 \quad \Rightarrow \quad &\, 
 -\mathrm{i}\,\frac{\bar{A}_{\sf w}}{(\bar{Y}^0)^2} \,\cR_{IJ}
\left( \frac{Y^I}{Y^0} - \frac{\bar{Y}^I}{\bar{Y}^0}\right)\,\left( \frac{Y^J}{Y^0} + \frac{\bar{Y}^J}{\bar{Y}^0}\right)\,,
\end{align}
where $A_{\sf w}$ is the scalar in the Weyl multiplet and the $Y^I$, $Y^0$ are standard vector multiplet projective coordinates, so that $z^I = Y^I/Y^0$, and
$\cK(Y ,\bar{Y})$ is the lowest order K\"ahler potential.

There are further invariants arising from the reduction, that cannot be currently described in
components\footnote{Note, however the final comments in section \ref{sec:H2F2}, which may lead to more general couplings.}
within supergravity, and are associated to terms involving $H^{2n}$ with $n$ odd, such as $H^2 F^2$, $H^2 R^2$ etc.
We comment on some of these terms, either giving the leading terms that characterise them, or pointing out their
apparent absence. 

An inventory of the four- and six-derivative couplings studied in this paper is given in Table \ref{tbl:sum}.
The first line in this table describes the terms based only on holomorphic functions of vector moduli and the
two chiral backgrounds. These are the only couplings that are controlled by a topological quantity, namely the vector of second Chern classes of the Calabi-Yau four-cycles. The gravitational $R^2$ coupling is the first nontrivial
coupling $F_{g} W^{2g}$ above, related to the topological string partition function
\cite{Bershadsky:1993cx, Antoniadis:1993ze}. The second, third and fourth lines correspond to the
non-holomorphic couplings of \cite{deWit:2010za}, where the tensor multiplet background is included. Note the
diagonal of underlined invariants of the type $R^2 F^{2n}$, which correspond to the first nontrivial couplings
$F_{g,n} W^{2g} V^{2n}$, for $g=1$, recently discussed in
\cite{Antoniadis:2010iq, Antoniadis:2013bja, Antoniadis:2013mna}.  The diagonal of the blue boxed invariants gives the one-loop copings of vector multiplets only, controlled by the tensors \eqref{eq:R-expr} and are the ones defining the structure of all other invariants in each line. To the best of our knowledge, the string original of such couplings have not been discussed in the literature.  The remaining invariants in the last line
can arise a priori and their description remains unknown within supergravity. We comment on the expected
structure of some of these below\footnote{In fact, we find that some of these couplings involving the tensor
multiplet seem to be missing in the specific compactification we consider, but cannot be excluded if more
tensor/hyper multiplets are considered.}.

\begin{table}[t]
\centering
{\renewcommand{\arraystretch}{1.7}
\begin{tabular}{|c|c|c|c|}
\hline
 \backslashbox{invariants}{derivatives}  & 4 \, & 6 \,& 8  \\ \hline
$ F(X, A_{\sf w}, A_{\sf t})$ & \underline{$R^2$},\,\,$(\nabla H)^2$ & -- & --  
\\ \hline
$\big[ F^2 + R^2 + (\nabla H)^2 \big]^2$ & \colorbox{LightBlue}{$(\nabla F)^2$} & 
       \underline{$R^2 F^2$},\, $(\nabla H)^2\, F^2$ & $R^4$,\, $R^2 (\nabla H)^2$,\, $(\nabla H)^4$  
\\ \hline
$\big[ F^2 + R^2 + (\nabla H)^2 \big]^3$ & -- &\colorbox{LightBlue}{$F^2(\nabla F)^2$} & \underline{$R^2 (\nabla F)^2$},\, $(\nabla H)^2 (\nabla F)^2$ 
\\ \hline
$\big[ F^2 + R^2 + (\nabla H)^2 \big]^4$ & -- & -- & \colorbox{LightBlue}{$(\nabla F)^4$} 
\\ \hline
Unknown & $H^2 F^2$ & $H^2 F^4$ & $R^4$,\, $H^6 F^2$,\, $H^2 F^6$  \\ \hline
\end{tabular}
}
\caption{A summary of higher-derivative couplings discussed here. The first row corresponds to chiral couplings
involving the Weyl and tensor multiplets. The next three rows display the known non-chiral $\cN\!=\!2$ invariants
at each order of derivatives, while the last row summarises the currently unknown invariants that can arise.
The double appearance of $R^4$ at the eight-derivative level corresponds to two different invariants
(see \eqref{eq:R4-red} below).  }
\label{tbl:sum}
\end{table}

\medskip
\noindent
The structure of the paper is as follows: In the next section we shall review briefly the one-loop $R^4$ couplings as well as some of our conventions and the reduction ansatze. The structure of known higher derivative couplings in four-dimensional $\cN\!=\!2$ theories is presented in sec. \ref{sec:n2d4}. We then proceed to consider the various higher derivative terms arising from the Calabi-Yau compactification of the one-loop term, organised by the order of derivatives. Hence, in section \ref{sec:eightder} we consider the eight derivative terms, while in sections \ref{sec:sixder} and \ref{sec:fourder} we discuss the six and four derivative invariants respectively. Some open questions are listed in section \ref{sec:open}. The extended appendices contain further technical details of the structures appearing in the main text. In particular, appendix \ref{sec_A:R4in10d} contains the fully explicit expressions for the quartic one-loop terms in 10D. Appendices \ref{App:4D-chiral-multiplets} 
and 
\ref{sec:tensor} deal with chiral 
couplings of general chiral multiplets and the composite chiral background of the tensor multiplet respectively. Finally, appendix \ref{app:kinetic} reviews the structure of the kinetic chiral multiplet and the various invariants that can be constructed based on it, up to the eight derivative level.

\section{Higher derivative terms in Type II theories}

The starting point for our considerations is the ten-dimensional eight-derivative terms that arise
in Type II string theories. The structure of the gravitational part of these couplings has been
known for a long time, but the coupling to the remaining Type II massless fields was not explicitly
known. Recently, a more concrete understanding of the terms involving the NS three-form field strength,
$H$, has been achieved \cite{Liu:2013dna}. The structure of the corresponding terms involving RR gauge
fields is constrained to a large extend, using arguments based on the eleven dimensional uplift to M-theory.

Upon reduction on a Calabi-Yau manifold without turning on any internal fluxes, the NS three form
leads to two types of objects in the four dimensional effective theory, namely a lower dimensional
three-form field strength and $h^{1,1}$ scalars. The former is naturally part of a tensor
multiplet\footnote{Upon  of the two-form gauge field to a scalar, this
leads to the so called universal hypermultiplet, but we will not consider this operation here.},
while the latter are part of vector multiplets in Type IIA and tensor/hyper multiplets in Type IIB.

In this section, we start by giving an overview of the ten-dimensional eight-derivative terms in
section \ref{sec:10-act}, from which all the lower dimensional higher-derivative terms arise. In
section \ref{sec:CY-red} we then turn to a discussion of the reduction procedure on Calabi-Yau
three-folds, which is central to the derivation of four dimensional couplings.

\subsection{The eight-derivative terms in ten dimensions}
\label{sec:10-act}

In summarising the structure of $R^4$ with the NS three-form $H$ included, it is most convenient to start by introducing the connection with torsion which reads in components
\begin{equation}
\Omega_{\pm \,\mu_1}{}^{\nu_1\nu_2}=\Omega_{\mu_1}{}^{\nu_1\nu_2} \pm \ft12H_{\mu_1}{}^{\nu_1\nu_2}.
\end{equation}
The curvature computed out of $\Omega_{\pm}$ is then
\begin{equation}
R(\Omega_{\pm})=R \pm \ft12d\mathcal H+\ft14\mathcal H\wedge\mathcal H,\qquad
\mathcal H^{\nu_1\nu_2}=H_{\mu_1}{}^{\nu_1\nu_2}dx^\mu.
\end{equation}
Denoting  the Riemann tensor by $R_{\mu\nu}{}^{\nu_1\nu_2}$ , we may write in components
\begin{equation}
R(\Omega_{\pm})_{\mu_1\mu_2}{}^{\nu_1\nu_2}=R_{\mu_1\mu_2}{}^{\nu_1\nu_2}
 \pm \nabla_{[\mu_1}H_{\mu_2]}{}^{\nu_1\nu_2}+\ft12H_{[\mu_1}{}^{\nu_1\nu_3}H_{\mu_2]\nu_3}{}^\beta.
\end{equation}
Note that the first and last term in this expression satisfy the pair exchange property,
while the second term is antisymmetric under pair exchange due to the Bianchi identity on
the three-form.
 
The Type II eight-derivative terms can be written in terms of two standard "$\mathcal{N}=1$
superinvariants", defined as  
\begin{align}
J_0(\Omega) = &\, \left(t_8 t_8 + \ft18\, \epsilon_{10} \epsilon_{10}\right)R^4 
\equiv \left(t_8 t_8 + \ft18 \epsilon_{10} \epsilon_{10}\right)^{\nu_1\dots\nu_8}_{\mu_1\dots\mu_8}
 R^{\mu_1 \mu_2}{}_{\nu_1\nu_2} \dots  R^{\mu_7 \mu_8}{}_{\nu_7\nu_8} \,,
\nonumber\\
J_1(\Omega) = &\, t_8 t_8 R^4  -  \frac14\, \epsilon_{10} t_8 B R^4 
 \equiv 
t_8 t_8 R^4 -  \frac14 \,t_8{}_{\mu_1\dots\mu_8} B\wedge R^{\mu_1 \mu_2} \wedge \dots  \wedge R^{\mu_7 \mu_8} 
 \,,
\label{eq:R4inv}
\end{align}
which provide a convenient way of encoding the kinematic structure of $R^4$ terms. The tensor $t_8$
and the associated tensorial structures appearing here are spelled out in appendix \ref{sec_A:R4in10d}.
Note that at this stage the terms \eqref{eq:R4inv} are build from Levi-Civita connections only,
and the three-from $H$ is not included. 

It has been argued in \cite{Liu:2013dna} that these will be completed  with the $B$-field as follows:
\begin{subequations}
\begin{eqnarray}
\label{superinv}
J_0(\Omega)  &\longrightarrow&
J_0 (\Omega_+) + \Delta J_0(\Omega_+, H)  \\
&& = \left(t_8 t_8 + \frac18 \epsilon_{10} \epsilon_{10}\right)R^4(\Omega_+)
+ \frac 13 \epsilon_{10} \epsilon_{10} H^2 R^3 (\Omega_+) + ...  \nn \\
J_1(\Omega)  &\longrightarrow&
J_1(\Omega_+) = 
t_8 t_8 R^4 (\Omega_+)  -  \frac18 \epsilon_{10} t_8 B \left(R^4 (\Omega_+) + R^4(\Omega_-) \right).
\label{superinv-b}
\end{eqnarray}
\end{subequations}
Note that $J_0 (\Omega_+) + \Delta J_0(\Omega_+, H)$ appears at tree level both in IIA and IIB
and at one loop in IIB, while $J_0 (\Omega_+) - 2 J_1(\Omega_+) + \Delta J_0(\Omega_+, H)$
appears at one loop in IIA. The  structure of $\Delta J_0(\Omega_+, H)$ is  more elaborate and kinematically different form the standard $\ft18 \epsilon_{10} \epsilon_{10} R^4 (\Omega) $ terms, and in fact it is the only part of the eight-derivative action that is not written purely in terms of $R(\Omega_{\pm})$.\footnote{Incidentally, using the connection with torsion $\Omega_{\pm}$ and $R(\Omega_{\pm})$ is not sufficient for writing the two-derivative effective action. For this one also needs the Dirac operator that appears in supersymmetry variations. Note that the Dirac operator, the covariant derivative with respect to $\Omega_{\pm}$ and the effective action are related via generalisation of the Lichnerowicz formula.} Here we should also use the full six-index un-contracted combination of $H^2$. These structures receive contributions starting form five-point odd-odd amplitudes:
\begin{align}
\label{}
 \Delta J_0(\Omega_+, H) = &\,
-\frac13\,\epsilon_{\alpha\mu_0\mu_1\cdots\mu_8}\epsilon^{\alpha\nu_0\nu_1\cdots\nu_8}
\,R^{\mu_7\mu_8}{}_{\nu_7\nu_8}(\Omega_+)
\nn\\
&\qquad
\times \big[ H^{\mu_1\mu_2}{}_{\nu_0}H_{\nu_1\nu_2}{}^{\mu_0}\,
   R^{\mu_3\mu_4}{}_{\nu_3\nu_4}(\Omega_+) R^{\mu_5\mu_6}{}_{\nu_5\nu_6}(\Omega_+)
\nn \\
&\qquad\quad
-\frac3{16}\,
(9\,H^{\mu_1\mu_2}{}_{\nu_0}H_{\nu_1\nu_2}{}^{\mu_0}+\ft19\, H^{\mu_1\mu_2\mu_0}H_{\nu_1\nu_2\nu_0}) \,
 \nabla^{\mu_3}H^{\mu_4}{}_{\nu_3\nu_4} \nabla^{\mu_5}H^{\mu_6}{}_{\nu_5\nu_6} \big]
\nn \\
&+ \ldots.
\label{eq:ooLaghat}
\end{align}
The order $H^4 R^2$ contribution is known up to some ambiguities, while the terms with higher
powers of $H$ remain a conjecture. Luckily these terms play little role in $\cN\!=\!2$ reductions and we comment on the cases where they are relevant below.

The last term in (\ref{superinv-b}), coming form the worldsheet odd-even and even-odd structures corresponds to the gravitational anomaly-canceling term. The relative sign between the two terms is fixed by the IIA GSO projection,
so that the coupling contains only odd powers of $B$-field. The explicit contribution to the effective action is 
\begin{gather}
\label{eq:x8}
- (2 \pi)^6 \alpha'^3  B \wedge \overline{X}_{8} = - \frac{ (2\pi)^2}{192}  \alpha'^3 B\wedge \left(\tr R^4-\frac14(\tr R^2)^2 + \mbox{exact } \right) \,,
\\
\overline{X}_{8} =\frac12 \, \left[ t_8 R^4 (\Omega_+) + t_8 R^4(\Omega_-) \right]\,. \nonumber
\end{gather}
Since $t_8 R^4 \sim \,  \frac14 p_1^2 - p_2$ is made of characteristic  classes and $H$ enters in (\ref{eq:x8}) like a torsion in the connection, its contribution amounts to a shift by exact terms. For completeness, we record the complete expression,
\begin{align}
\label{eq:x8shift}
\overline{X}_{8} =&\, \frac1{ 192 (2\pi)^4}\bigg[  \left( \tr R^4 -\frac1{4}(\tr R^2)^2 \right) 
\nn \\
\hspace{1cm} &\quad 
+ d \,\, \bigg( \frac1{2} \tr \left( \HH \nabla \HH R^2 + \HH R \nabla \HH R
+ \HH R^2 \nabla \HH \right) 
  \nn \\
& \qquad\qquad 
-\frac1{8} \left( \tr R^2 \, \tr \HH \nabla \HH + 2\, \tr \HH R \, \tr R \nabla \HH \right)  
  \nn \\
& \qquad\qquad +\frac1{16}\,  \tr  \left( 2 \HH^3 (\nabla \HH R + R \nabla \HH) + \HH R \HH^2 \nabla \HH
+ \HH \nabla \HH \HH^2 R \right)\nn\\
& \qquad\qquad - \frac1{16}\,\left( \tr \HH \nabla \HH \,  \tr R \HH^2 +  \tr R \nabla \HH \,  \tr \HH^3
- \tr \nabla \HH \HH^2 \, \tr \HH R \right) 
 \nn \\
& \qquad\qquad +\frac1{32} \tr \nabla \HH \HH^5 + \frac1{16}  \tr \HH (\nabla \HH)^3  
 \nn \\
& \qquad\qquad + \frac1{192}  \tr \nabla \HH \HH^2 \, \tr \HH^3 
  -\frac1{64} \tr \HH \nabla \HH \, \tr (\nabla \HH)^2  \bigg) \bigg].
\end{align}
since its reduction will be useful in the following.

The eight-derivative (tree level and one-loop) terms are the origin of the only perturbative corrections to the metrics on the $\cN\!=\!2$ moduli spaces. The corrections respect the factorisation of the moduli spaces, and the classical metrics on moduli space of vectors and hypers receive respectively tree-level and one-loop corrections, both of which are proportional of the Euler number of the internal Calabi-Yau manifold  \cite{Antoniadis:1997eg, Antoniadis:2003sw}.  Needless to say, our discussion is consistent with these corrections, and from now on we shall concentrate only on the higher-derivatives terms. Recent progress in understanding the hyper-multiplet quantum corrections  is reviewed in \cite{Alexandrov:2013yva}.

As already mentioned,  the reduction of type IIA super invariant $J_0 (\Omega) - 2 J_1(\Omega)$ on Calabi-Yau manifolds yields the one loop $R^2$ terms in $\cN\!=\!2$ four-dimensional theory, and this is the only known product of the reduction so far that leads to higher derivative terms in 4D. We shall return to the four-dimensional $R^2$ terms in section \ref{sec:fourder}. Clearly, the inclusion of the $B$ field leads to further couplings to matter upon dimensional reduction, to which we now turn.

\subsection{Reduction on Calabi-Yau manifolds}
\label{sec:CY-red}

We now consider the reduction of the ten-dimensional eight-derivative action on a Calabi-Yau threefold $X$, and its relation to the $\cN\!=\!2$ action. The metric can be reduced in the standard way, as\footnote{We use numbered Greek letters for 10D curved indices, while ordinary Greek letters denote 4D curved indices. We use Latin letters from the beginning of the alphabet for 4D flat indices, and Latin letters from the middle to the end of the alphabet, $m, n, \dots$ are reserved for CY indices. We reserve the letters $i,j,k,l$ for $SU(2)$ R-symmetry indices. Capital Latin indices $I,J = 1, ..., h^{1,1}(X)$ span the matter vector multiplets.}
\begin{equation}
 g_{\mu_1\mu_2} = \begin{pmatrix}  g_{\mu\nu} & 0 \\   0   &   g_{mn}   \end{pmatrix}\,,
\end{equation}
where $g_{mn}$ is the metric on the Calabi-Yau manifold, X, which we will not need explicitly. The three-form $H$ reduces as
\begin{eqnarray}
H_3 = H + f^I \wedge \omega_I\,,
\label{eq:H3red}
\end{eqnarray}
where the four-dimensional $H$ is part of the tensor multiplet, and the one-forms $f^I$ can be locally written as $f^I = d u^I$, with $u^I$ being a part of the vector multiplet scalars. The index $I$ spans over $h^{1,1}(X)$, and $\omega_I \in H^2(X, \mathbb{Z})$. 

Hence reducing the terms built out of $R(\Omega_{\pm})$ and $H_3$  (where $\Omega_{\pm}=\Omega \pm\ft12\HH$), one expects at a given order of derivatives various couplings involving the Riemann tensor, $R$, as well as the tensor multiplet and vector multiplets. For example, at the four-derivative level one recovers the four-dimensional $R^2$ couplings and expects to obtain further couplings quartic in tensor multiplet and vector multiplets, as well as mixed terms. We use the symbolic computer algebra system Cadabra \cite{DBLP:journals/corr/abs-cs-0608005, Peeters:2007wn} to systematically derive the structure of these terms.

The vector moduli shall be denoted $z^I = u^I + i t^I$, where $t^I$ are the K\"ahler moduli, defined through the decomposition of the Calabi-Yau K\"ahler form, $J$, as $J = t^I \omega_I$. The total volume, $\cV$, of the CY manifold is given by the standard volume form, cubic in $J$ as
\begin{equation}
 \cV= \frac{1}{3!}\,\int_X J\wedge J\wedge J = \log[-\cK]\,,
\end{equation}
where we defined the 4D K\"ahler potential. We shall not need the vector fields themselves, but only their field strengths, denoted as $F^A$, where the index $A={0,I}$ runs over the $h^{1,1}(X)+1$ vector fields (in places where the shorthand notation is used, $F$ will stand for the entire multiplet). 

Reducing the NS eight-derivative couplings we obtain couplings that contain $u^I$ and $t^I$. The couplings to $F^I$ can be recovered by thinking of the (one-loop) couplings as being reduced from five (or eleven) dimensions. In practical terms, one has to add an extra index on $f^I_{\mu} \, \mapsto F^I_{\mu \nu}$. Since (the affected parts of) the expressions are even in powers of $F^I$, the extra index will always be contracted with a similar counterpart. Moreover most of the expressions are only quadratic in $F$, hence the lifting is unique. A little combinatorial imagination is needed for $F^4$ terms. This procedure follows the lifting of one-loop NS couplings to eleven dimensions, outlined in \cite{Liu:2013dna} and is analogous to the way one can recover graviphoton couplings from the $R^2$ term - one just has to think of the lifting of the couplings to five dimensions and their consequent reduction.
As already mentioned, here we can benefit from the explicit $\cN\!=\!2$ formalism in verifying that the couplings involving $t^I$ and $F^I$ complete each other sypersymmetrically and hence provide a verification of the lifting of the complete one-loop eight-derivative terms from type IIA strings to M-theory.

Since we are focusing on Calabi-Yau compactifications without flux, different pieces in the reduction will involve integrating over $X$  expressions  containing some power of the internal curvature and $\omega_I \in H^2(X, \mathbb{Z})$.\footnote{Since the four-dimensional three-form $H$ in \eqref{eq:H3red} is in the hyper matter, some of the couplings involving hyper multiplets will be discussed here. However we mostly  concentrate on the vector multiplets here, and do not consider any internal expressions involving forms in $H^{2,1}(X)$.} We shall start with the familiar integrals.

At the four derivative level, one needs to consider terms with exactly two powers of the Riemann tensor
in the internal Calabi-Yau manifold. In the purely gravitational sector, one then finds an $R^2$ term
in four dimensions, originating from the $R^4$ couplings in ten dimensions. In this case, one obtains
\begin{equation}
 t_8 t_8 R^4 = -\frac18\epsilon_{10} \epsilon_{10} R^4 =
12\, F_1\, R^{\mu\nu\rho\lambda}R_{\mu\nu\rho\lambda} \,,
\end{equation}
where we note that only terms completely factorised in internal and external objects contribute.
The function $F_1$ is an integral over the internal directions that takes the form
\begin{equation}\label{eq:zero-id}
 F_1 = \int_{X} R^{mn pq}R_{mn pq}
= \frac{1}{8}\,\int_{X} \epsilon_{mn m_1 \ldots m_4} \epsilon^{mn n_1 \ldots n_4}
 R^{m_1 m_2}{}_{n_1n_2} R^{m_3 m_4}{}_{n_3n_4}
= \alpha_I t^I\,,
\end{equation}
where the first equality holds up to Ricci terms and in the second equality we evaluated the integral.

The fine balance between the two a priori different terms in \eqref{eq:zero-id} can be extended
to more complicated integrals, that are relevant in the reduction of the non-purely gravitational
terms. In this case, we have checked explicitly the identity
\begin{align}\label{eq:two-id}
&\,t_8{}_{\mu m \nu n m_1 \ldots m_4} t_8{}^{\rho p \sigma q n_1 \ldots n_4}
\omega_I{}^m{}_p \omega_J{}^n{}_q \, R^{m_1m_2}{}_{n_1n_2} R^{m_3 m_4}{}_{n_3n_4}
=
\nonumber\\
&\,\hspace{4.5cm}
-\frac{1}{8}\, \delta_{\mu}{}^{\sigma}\, \delta_{\nu}{}^{\rho}
\epsilon_{mn m_1\ldots m_4} \epsilon^{pq n_1\ldots n_4}
\omega_I{}^m{}_p \omega_J{}^n{}_q \, R^{m_1m_2}{}_{n_1n_2} R^{m_3 m_4}{}_{n_3n_4}\,,
\end{align}
up to Ricci terms. Note that the structure of spacetime indices is different in the two
sides, while the remaining terms are purely internal.
Upon contraction with the K\"ahler moduli, each side of \eqref{eq:two-id} reduces to
the expression in \eqref{eq:zero-id}.

Even further, the identity in \eqref{eq:four-id} can be generalised to an identity
involving eight indices, as follows.
\begin{align}\label{eq:four-id}
&\,t_8{}_{m_1\dots m_4 p_1 \ldots p_4} t_8{}^{n_1\dots n_4 q_1 \ldots q_4}
\omega_I{}^{m_1}{}_{n_1} \omega_J{}^{m_2}{}_{n_2} \, \omega_K{}^{m_3}{}_{n_3} \omega_L{}^{m_4}{}_{n_4} \,
R^{p_1p_2}{}_{q_1q_2} R^{p_3 p_4}{}_{q_3q_4}
=
\nonumber\\
&\,\hspace{1.5cm}
-\frac{1}{8}\,
\epsilon_{p_0 q_0 m_1\dots m_4 p_1\ldots p_4} \epsilon^{p_0 q_0 n_1\dots n_4 q_1\ldots q_4}
\omega_I{}^{m_1}{}_{n_1} \omega_J{}^{m_2}{}_{n_2} \, \omega_K{}^{m_3}{}_{n_3} \omega_L{}^{m_4}{}_{n_4} \,
R^{p_1p_2}{}_{q_1q_2} R^{p_3 p_4}{}_{q_3q_4}\,,
\end{align}
again up to Ricci-like terms. These two expressions are relevant for the terms in $R(\Omega_+)$
that are odd or even under pair exchange, respectively.

The reduction to six- and eight-derivative couplings will require integration over expressions
linear or zeroth order in the  Riemann tensor of the internal Calabi-Yau manifold. In view of
the vanishing of the Calabi-Yau Ricci tensor, these are essentially unique, and given by
\begin{align}
 G_{IJ} =&\, \tfrac12\, \int_X\!\omega_I^{mn} \omega_J{}_{mn}\,,
\nonumber\\
\cR_{IJ} =&\, \int_{X} R_{mnpq}\,\omega_{I}{}^{mn}\omega_{J}{}^{pq}\,,
\end{align}
where $G_{IJ}$ ultimately leads to the vector multiplet K\"ahler metric and $\cR_{IJ}$ is
a new coupling to be discussed in due time.

\section{The four dimensional action}
\label{sec:n2d4}

We now describe the structure of the effective $\cN\!=\!2$ supergravity action in four dimensions,
that arises from the reduction of the one-loop Type IIA Lagrangian. Given that the original
ten dimensional action contains eight derivatives, one obtains a variety of
higher derivative couplings, next to the lowest order two derivative action.
In order to describe these in a systematic way, we will consider the off-shell formulation
of the theory, which allows to construct infinite classes of higher derivative invariants
without modifying the supersymmetry transformation rules. However, since the higher
dimensional one-loop action and the reduction scheme are on-shell, one has to deduce the
off-shell invariants from the desired terms that result upon gauge fixing to the on-shell
theory. In the following, we take the pragmatic approach of matching the leading,
characteristic terms in each invariant and promoting to off-shell variables by standard
formulae for special coordinates for the vector multiplet scalars. In practice, these
choices are essentially unique, and below we comment on this issue in the examples where
this is relevant.

The defining multiplet of off-shell $\cN\!=\!2$ supergravity is the Weyl multiplet, which
contains the graviton, $e^a_\mu$, the gravitini, gauge fields for local scale and R-symmetries
and various auxiliary fields. Of the latter, only the auxiliary tensor $T_{ab}{}^{ij}$ is
directly relevant, since it is identified with the graviphoton in the on-shell formulation
of the theory, at the two-derivative level. The reader can find a short account of the Weyl
multiplet in Appendix \ref{App:N2sugra}.
In what follows, we will mostly deal with the covariant fields of the Weyl multiplet,
which can be arranged in a so-called chiral multiplet (see Appendix \ref{App:4D-chiral-multiplets}
for more details), which contains the auxiliary tensor $T_{ab}{}^{ij}$
and the curvature $R(M)_{\mu \nu}{}^{ab}$. The latter is identified with the Weyl tensor,
up to additional modifications. These observations will be very useful in the identification
of the various higher derivative couplings.

There are various matter multiplets that can be defined on a general
supergravity background. Here, the fundamental matter multiplets we consider are
vector multiplets and a single tensor multiplet, corresponding to the universal
tensor multiplet of Type II theories. Both these multiplets comprise $8 + 8$
degrees of freedom and are defined in appendices \ref{App:4D-chiral-multiplets}
and \ref{sec:tensor} respectively, to which we refer for further details. 
Moreover, they can be naturally viewed as two mutually non-compatible projections of
a chiral multiplet, which is central to our considerations.

All Lagrangians considered in this paper are based on couplings of chiral multiplets, which
contain $16 + 16$ degrees of freedom and can be defined on an arbitrary
$\cN\!=\!2$ superconformal background. We refer to appendix \ref{App:4D-chiral-multiplets} for
more details on chiral multiplets. Here, we simply state that these multiplets are labeled
by the scaling weight, $w$, of their lowest component, $A$, and that products of chiral
multiplets are chiral multiplets themselves, obtained by simply considering functions
$F(A)$, which must be homogeneous, so that a weight can be assigned to them. As mentioned
above, the matter multiplets we consider are also chiral multiplets of $w=1$, on which a
constraint projecting out half of the degrees of freedom is imposed and the same property
holds for the covariant components of the Weyl multiplet. This implies that
actions for all the above multiplets can be generated by considering expressions
constructed out of chiral multiplets, which are invariant under supersymmetry.

\subsection{Two derivatives}

The prime example is given by the invariant based on a $w=2$ chiral multiplet, implying
that its highest component, $C$, has Weyl weight 4, and chiral weight 0, as is appropriate
for a conformally invariant Lagrangian in four dimensions. It can be shown that the expression
\begin{align}
  \label{eq:chiral-density}
  e^{-1}\mathcal{L} =&\, C -\tfrac1{16}A( T_{ab\,ij} \varepsilon^{ij})^2 
  + \text{fermions}\,,
\end{align}
is the bosonic part of the invariant, including a conformal supergravity background described by
the auxiliary tensor $T_{ab\,ij}$ of the gravity multiplet.
The two derivative action for vector multiplets is now easily constructed, by setting the chiral
multiplet in this formula to be composite, expressed in terms of vector multiplets labeled by
indices $I,J,\dots= 0,1,\ldots,\nv$. It is possible to show (cf.~\eqref{eq:chiral-mult-exp})
that the relevant terms of such a composite multiplet are given by\footnote{The function $G$
in \eqref{eq:chiral-mult-exp} is conventionally chosen as $G(X^I)= -\tfrac {\mathrm{i}} 2 F(X)$
in this context.}
\begin{align}
 \label{eq:chiral-mult-comp}
  A =&\, -\tfrac {\mathrm{i}} 2 F(X) \,,\nonumber\\
  C =&\, \mathrm{i}\,F(X)_I\, \Box_\mathrm{c}  \bar X^I  
  +\tfrac{\mathrm{i}}8\, F(X)_{IJ}\big[ B_{ij}{}^I B_{kl}{}^J\,
   \varepsilon^{ik} \varepsilon^{jl} + X^I\,G^{+}_{ab}{}^J T^{ab}{}_{ij} \varepsilon^{ij}
   -2\, G^{-}_{ab}{}^I G^{-abJ}\big]  \,,
\end{align}
where $F_{I}$ and $F_{IJ}$ are the first and second derivative of the function $F$, known
as the prepotential and $B_{ij}{}^I$, $G^{-}_{ab}{}^I$ are the remaining bosonic components
of the chiral multiplets (which in this case are constrained by \eqref{eq:vect-mult} for
vector multiplets). As the bottom composite component, $A$, has $w=2$, the function $F(X)$
must be homogeneous of degree two in the vector multiplet scalars $X^I$. Taking into account
the constraints in \eqref{eq:vect-mult}, the bosonic terms of the Lagrangian following from
\eqref{eq:chiral-density} read
\begin{eqnarray}\label{eq:4d-lagr-v}
8\pi\,e^{-1}\, {\cal L}_{v} &=&
 \mathrm{i} {\cal D}^{\mu} F_I \, {\cal D}_{\mu} \bar X^I   - \mathrm{i} F_I\,\bar X^I
 (\ft16  R - D)
-\ft18\mathrm{i}  F_{IJ}\, Y^I_{ij} Y^{Jij}  \nonumber\\
&&+\ft14 \mathrm{i} F_{IJ} (F^{-I}_{ab} -\ft 14 \bar X^I
T_{ab}^{ij}\varepsilon_{ij})(F^{-Jab} -\ft14 \bar X^J
T^{ijab}\varepsilon_{ij})  \nonumber\\
&&-\ft18 \mathrm{i} F_I(F^{+I}_{ab} -\ft14  X^I
T_{abij}\varepsilon^{ij}) T^{ab}_{ij}\varepsilon^{ij}
-\ft1{32} \mathrm{i} F (T_{abij}\varepsilon^{ij})^2 + {\rm h.c.}\;,
\end{eqnarray}
where in the last line we added the hermitian conjugate to obtain a real Lagrangian.
Here, $F^{I}_{ab}$ are the vector multiplet gauge field strengths, $R$ is the Ricci
scalar and $D$ is the auxiliary real scalar in the gravity multiplet.
This Lagrangian is invariant under scale transformations and can be related to an
on-shell Poincar\'e Lagrangian by using a scale transformation to set the coefficient
of the Einstein-Hilbert term, $\I (F_I\,\bar X^I)$, to a constant. For standard Calabi-Yau
compactifications of Type II theories, one obtains a cubic prepotential, as
\begin{equation}\label{eq:CY-prep}
 F = -\frac16\,\frac{C_{IJK}Y^I Y^J Y^K }{Y^0}\,, 
\end{equation}
where the constant tensor $C_{IJK}$ stands for the intersection numbers of the manifold.

As it turns out, the Lagrangian \eqref{eq:4d-lagr-v} is inconsistent as it stands, so
that one needs to add at least one auxiliary hypermultiplet, which is to be gauged away
by superconformal symmetries, similar to the scalar $\I (F_I\,\bar X^I)$ above.
In addition, in this paper we consider a single tensor multiplet, corresponding to
the universal hypermultiplet upon dualisation of the tensor field. We refer to
appendix \ref{sec:tensor} for more details on this multiplet. For later reference, we
display the bosonic action for the auxiliary hypermultiplet and the physical tensor multiplet
that needs to be added to \eqref{eq:4d-lagr-v} to obtain a consistent on-shell
theory with a physical tensor multiplet, as
\begin{align}\label{eq:4d-lagr-t}
8\pi\,e^{-1}\, {\cal L}_{t} =&
- \ft12 \varepsilon^{ij}\,\Omega_{\a\b} \,\big[
{\cal D}_\mu A_i{}^\a \,{\cal D}^\mu  A_j{}^\b 
- A_i{}^\alpha A_j{}^\beta \big(\tfrac{1}{6} R  + \tfrac12\, D)  \big]
\nonumber\\
 &\,
 - \, \tfrac{1}{2} \, F^{(2)} \,\mathcal{D}_\mu L_{ij} \, \mathcal{D}^\mu L^{ij}
 + F^{(2)} \, L_{ij} \, L^{ij} \, \Big(\tfrac{1}{3} R +  D \Big)
 + F^{(2)} \, \Big[ E_{\mu} \, E^{\mu} + G \bar{G} \Big]
\nonumber\\
 &\, + \tfrac12 \mathrm{i} e^{-1}
   \varepsilon^{\mu \nu \rho \sigma} \, \frac{\partial F^{(2)}}{\partial L_{ij}}\,E_{\mu\nu} \,
   \, \partial_\rho L_{ik} \, \partial_\sigma L_{jl} \,\varepsilon^{kl}  \, .
\end{align}
Here, $A_i{}^\a$ is the hypermultiplet scalar, described as a local section of
$\mathrm{SU}(2)\times\mathrm{SU}(2)$, and $\Omega_{\alpha\beta}$ is the invariant
antisymmetric tensor in the second $\mathrm{SU}(2)$.
The on-shell fields of the tensor multiplet are the triplet of scalars $L_{ij}$ and the 
two-form gauge field, $B_{\mu\nu}$, while 
\begin{equation}
E^\mu = \tfrac{1}{2}\mathrm{i}\, e^{-1} \, \varepsilon^{\mu \nu
\rho \sigma} \partial_\nu B_{\rho \sigma}\,, 
\end{equation}
is the dual of its field strength and $G$ is a complex auxiliary scalar. The
functions $F^{(2)}$ and $F^{(3)}$ can be viewed as the second and third
derivative of a function of the $L_{ij}$, that can easily be generalised to
an arbitrary number of tensor multiplets \cite{deWit:2006gn}
(see \eqref{eq:chiral-constraints}-\eqref{eq:sc-tensor}). For a single tensor
multiplet, there is a unique choice, as 
\begin{align}\label{eq:two-der-ten}
 F^{(2)} = \frac{1}{\sqrt{L_{ij} L^{ij} } }\,,
\end{align}
which we will assume throughout. However, as we ignore all tensor multiplet scalars in
our reduction scheme, all scalars and $F^{(2)}$ are kept constant and only appear as overall
factors.

\subsection{Higher derivatives}

In this paper we construct higher derivative actions based on the properties of chiral
multiplets, as discussed above. One way of doing this is to consider the function
$F$ in \eqref{eq:chiral-mult-comp} to depend not only on vector multiplets, but also on
other chiral multiplets, which are treated as background fields. Alternatively, one
may consider invariants more general than \eqref{eq:chiral-density}, containing explicit
derivatives on the chiral multiplet fields. Here we use both structures, which we
discuss in turn, emphasising the methods and the structure of invariants rather than
details, which can be found in \cite{de Wit:1996ix, deWit:2006gn, deWit:2010za}.

We consider two chiral background multiplets, one constructed out of the Weyl multiplet and one
constructed out of the tensor multiplet, whose lowest components we denote as
$A_{\sf w}$ and $A_{\sf t}$ respectively. These are proportional to the auxiliary fields
$(T_{ab}{}^{ij}\varepsilon_{ij})^2$ of the Weyl and and $G$ of the tensor multiplet and we
refer to appendices \ref{App:4D-chiral-multiplets} and \ref{sec:tensor} for more
details on their precise definition. Considering a function $F(X^I, A_{\sf w}, A_{\sf t})$
leads to a Lagrangian of the form \eqref{eq:4d-lagr-v}, where the set of
vector field strengths is extended to include the Weyl tensor $R(M)_{ab}{}^{cd}$
in \eqref{eq:curvatures-4} and the combination $\nabla_{[a} E_{b]}$, so that four
derivative interactions of the type 
\begin{equation}\label{eq:chiral-coup}
\mathcal{L} = \int F(X^I, A_{\sf w}, A_{\sf t}) \propto 
\int \left( \frac{\partial F}{\partial A_{\sf w}}\, R(M)^{-\,2} 
   + \frac{\partial F}{\partial A_{\sf t}}\,(\nabla_{[a} E_{b]_{-}})^2  + \dots \right)\,, 
\end{equation}
are generated.
The explicit expressions for the relevant chiral multiplets can be found in
\eqref{eq:W-squared} and \eqref{eq:CT} respectively.

These couplings are distinguished, in the sense that they are described by
a holomorphic function and correspond to integrals over half of superspace.
The $R^2$ term has been studied in detail, especially in connection to BPS black holes,
see e.g. \cite{Antoniadis:1997eg, Maldacena:1997de, Mohaupt:2000mj, LopesCardoso:2000qm, Banerjee:2011ts}.
The full function $F(X^I, A_{\sf w})$ is in this case related to the topological string
partition function \cite{Bershadsky:1993cx, Antoniadis:1993ze}. We will only be concerned
with the linear part of this function, originating in the one-loop term in section \ref{sec:10-act},
which is controlling the $R^2$ coupling through \eqref{eq:chiral-coup}. 
The $(\nabla E)^2$ term has appeared more recently \cite{deWit:2006gn},
without any coupling to vector multiplets.

More general higher derivative couplings can be constructed by looking for invariants
of chiral multiplets that contain explicit derivatives, unlike \eqref{eq:chiral-density}.
Indeed, such invariants can be derived by considering a chiral multiplet whose components
are propagating fields, i.e. described by a Lagrangian containing derivatives. This can
be done in the standard way, by writing a K\"ahler sigma model, which in the simple case
of two multiplets reads
\begin{equation}
  \int \!\mathrm{d}^4\theta\;\mathrm{d}^4\bar\theta \;\Phi\,\bar\Phi^\prime
  \approx
  \int \!\mathrm{d}^4\theta \,\Phi\,\mathbb{T}(\bar\Phi^\prime) \,,
\end{equation}
where both $\Phi$ and $\Phi^\prime$ must have $w=0$ for the integral to be well defined.
In the second form of the integral we defined a new chiral multiplet, $\mathbb{T}(\bar\Phi^\prime)$,
the so called kinetic multiplet, since it contains the kinetic terms for the various
fields. This multiplet was constructed explicitly in \cite{deWit:2010za} and is
summarized in appendix \ref{app:kinetic} below (see also \cite{Butter:2013lta} for a
recent generalisation). In practice, one can think of the operator $\mathbb{T}$ as an
operator similar to the Laplacian, acting on the components of the multiplet, as we find
\begin{align}
  \label{eq:quad-chir}
  \int \!\mathrm{d}^4\theta \,\Phi\,\mathbb{T}(\bar\Phi^\prime) =&\,
  C\,\bar C^\prime + 8\, \mathcal{D}_a F^{-ab}\, \mathcal{D}^c F^{\prime +}{}_{cb} 
  +4\,\mathcal{D}^2 A\,\mathcal{D}^2\bar A^\prime
  + \cdots \,, 
\end{align}
where we only display the leading terms.

One can now simply declare the chiral multiplets $\Phi$, $\Phi^\prime$ to be composite
by imposing \eqref{eq:chiral-mult-comp}, where the corresponding functions $F$, $F^\prime$
can depend on vector multiplet scalars, as well as the Weyl and tensor multiplet
backgrounds, exactly as described above. As described in section \ref{app:kinetic} and
in \cite{deWit:2010za}, this leads to a real function $\cH=F \bar F^\prime + \text{c.c.}$,
homogeneous of degree zero, which naturally describes a variety of higher derivative couplings,
corresponding to the combinations generated by
\begin{equation}
 \big[ F^{-\,2} + R^{-\,2} + (\nabla E)^{-\,2} \big] \otimes \big[ F^{+\,2} + R^{+\,2} + (\nabla E)^{+\,2} \big]\,,
\end{equation}
where the $\pm$ stand for selfdual and anti-selfdual parts.
Each of these is controlled by a function of the vector multiplet moduli as
\begin{align}
\label{eq:quad-expand}
\cH(X^I, A_{\sf w}, A_{\sf t}, \bar{X}^I, \bar{A}_{\sf w}, \bar{A}_{\sf t})= &\,
\sum_{i+j \leq 2} \cH^{(i,j)}(X^I, \bar{X}^I) (A_{\sf w})^i\, (\bar{A}_{\sf t})^j  + \text{c.c.}
\Rightarrow
\nonumber\\
\qquad \qquad 
 \cH^{(0,0)}(X^I, \bar{X}^I) &\,\qquad \Rightarrow \qquad (\nabla F)^2, F^4
\nonumber\\
 \qquad \qquad 
 \big[\cH^{(1,0)} A_{\sf w} + \text{c.c.}\big]  &\,\qquad \Rightarrow \qquad R^2 F^2
\nonumber\\
 \big[\cH^{(0,1)}\, A_{\sf t} + \text{c.c.}\big] &\,\qquad \Rightarrow \qquad (\nabla E)^2 F^2
\nonumber\\
 \qquad \qquad 
 \cH^{(2,0)} A_{\sf w}\, \bar{A}_{\sf w} &\,\qquad \Rightarrow \qquad R^4
\nonumber\\
 \cH^{(0,2)}\, A_{\sf t}\, \bar{A}_{\sf t} &\,\qquad \Rightarrow \qquad (\nabla E)^4
\nonumber\\
 \big[\cH^{(1,1)}\, A_{\sf w}\, \bar{A}_{\sf t} + \text{c.c.}\big]  &\,\qquad \Rightarrow \qquad R^2 (\nabla E)^2\,,
\end{align}
where we display the characteristic terms at each order.
Note that we consider a function at most quadratic in $A_{\sf w}$, $A_{\sf t}$, since
a higher polynomial would lead to the same terms, multiplied by additional powers of these
auxiliary scalars. These are analogous to the non-linear parts of the chiral coupling
in \eqref{eq:chiral-coup} and go beyond one-loop terms, so we do not consider them in
the following. Finally, note that due to the expansion \eqref{eq:quad-expand}, the functions $\cH^{(i,j)}$
are not homogeneous of degree zero for $i,j\neq 0$, but we we will always refer to the
corresponding degree zero monomial in \eqref{eq:quad-expand}, for clarity.

The invariants based on \eqref{eq:quad-chir} are the simplest ones containing
the kinetic multiplet. It is straightforward to construct more general integrals, for
example 
\begin{equation}
 \int \!\mathrm{d}^4\theta \,\Phi\,\mathbb{T}(\bar\Phi)\,\mathbb{T}(\bar\Phi) \,,
\quad
 \int \!\mathrm{d}^4\theta \,\Phi\,\mathbb{T}(\bar\Phi)\,\mathbb{T}(\bar\Phi)\,\mathbb{T}(\bar\Phi) \,,
\quad
 \int \!\mathrm{d}^4\theta \,\Phi_0\mathbb{T}(\bar\Phi)\,\mathbb{T}(\bar\Phi_0\mathbb{T}(\Phi)) \,,
\end{equation}
which are the cubic and quartic invariants discussed in section \ref{app:kinetic}. In exactly the
same way as above, the first of these integrals leads to a homogeneous function of degree
$-2$, describing couplings cubic in $F^2$, $R^2$ and $(\nabla E)^2$. Only some of these
are relevant in the following, in particular the $(R^2 +(\nabla E)^2)F^4$ and $F^6$, since
the rest contain more than eight derivatives. Finally, the last two integrals describe
couplings with at least eight derivatives and lead to homogeneous functions of degree $-4$.
Only the last integral is relevant for us, namely for the $F^8$ term.

\section{Eight derivative couplings}
\label{sec:eightder}

We start by considering terms containing the maximum number of derivatives appearing in
the one-loop correction, i.e. we consider the possible eight derivative invariants in
four dimensions. This may seem counterintuitive at first and in fact some of these
invariants have not been described explicitly. However, the terms that
are known in four dimensions are the simplest to describe, setting the stage for the more
complicated structures to follow. 

Applying the rules and assumptions spelled out in section \ref{sec:CY-red}, one can characterise
the various terms appearing in the reduction by the order of Riemann tensors, tensor multiplet
fields strengths and vector multiplet fields strengths arising in four dimensions. 
Schematically, we then find a decomposition of the type
\begin{align}\label{eq:8-dec}
\mathcal{L}^{1-\text{loop}} \Rightarrow &\, 
\textcolor{blue}{R^4} + \textcolor{blue}{R^2 (\nabla H)^2} + \textcolor{blue}{(\nabla H)^4} 
+ \textcolor{red}{ \underline{R^4}} + \textcolor{red}{\underline{H^6 F^2}} 
 + \textcolor{blue}{H^4 F^4} \nonumber\\
&\, 
+ \textcolor{red}{\underline{H^2 F^6}} 
 + \textcolor{blue}{R^2\,(\nabla F)^2} + \textcolor{blue}{(\nabla F)^4}\,,
\end{align}
where we write in blue the terms which correspond to the known four-dimensional  invariants. The supersymmetric invariants for the underlined (red) terms are not known.

\subsection*{Gravity and tensor couplings}
The most obvious and simplest term is the $R^4$ term, which arises by trivial reduction of the
corresponding ten dimensional term. Note that only the even-even contribution survives the
reduction and leads to a four dimensional $R^4$ term as
\begin{align}\label{eq:R4-red}
 t_8t_8 R^4 \rightarrow &\,
192\,(R_{\mu\nu \rho\lambda} R^{\mu\nu \rho \lambda})^2
+ 144\,\tr[R_{\mu\nu} R_{\rho\lambda}] \tr[R^{\mu\rho} R^{\nu\lambda}] + \dots
\nonumber\\
=&\,
768\,(R^+)^2 (R^-)^2
+48\,\left( (R^+)^2 -  (R^-)^2\right)^2 + \dots\,.
\end{align}
The second line corresponds to two different invariants in four dimensions, each
with its own supersymmetric completion, corresponding to the double appearance of $R^4$
in \eqref{eq:8-dec}. The supersymmetrisation of the second
term is not known in $\cN\!=\!2$ supergravity (see however \cite{Moura:2007ks} for a
discussion in the $\cN\!=\!1$ setting).
The supersymmetric completion of the first term was found in \cite{deWit:2010za},
where it was shown that it is governed by a homogeneous degree zero real function of
the vector multiplet moduli and the Weyl multiplet scalar $A_{\sf w}$. In the present
case however, \eqref{eq:R4-red} does not depend on moduli other than the total volume of
the CY manifold, so that we can immediately identify the relevant function as depending
only on the K\"ahler potential as
\begin{equation}\label{eq:R4-fun}
 \cH_{R^4}= \frac{3}{16}\,\mathrm{e}^{2\,\cK(Y ,\bar{Y})}\, A_{\sf w} \bar{A}_{\sf w} \,.
\end{equation}
Here, the function of the off-shell scalars $\cK(Y ,\bar{Y})$ is very closely related to
the lowest order K\"ahler potential, as
\begin{equation}
 \mathrm{e}^{-\cK} = 2\,\I F_{AB} Y^A\bar{Y}^B\,, 
\end{equation}
with the prepotential \eqref{eq:CY-prep}, and is equal to it once special coordinates
are chosen (for $Y^0=1$). Note however, that
this is only the most natural choice that results in the first coupling in \eqref{eq:R4-red}
upon taking the on-shell limit and one might consider more elaborate off-shell functions
leading to the same result.
Upon taking derivatives of this function with respect to the vector multiplet moduli,
various couplings involving vector multiplet field strengths and auxiliary fields 
arise at the off-shell level, resulting to further eight derivative terms in the on-shell
theory.

The corresponding purely tensor coupling is the eight derivative term of the tensor
multiplet, which takes the form
\begin{equation}
 \label{eq:H8-red}
 t_8t_8 R^4 \rightarrow 
96\,\left( (\nabla_{[\mu} E_{\nu]} \nabla^{[\mu} E^{\nu]})^2 
- 4\,\nabla_{[\mu} E_{\nu]} \nabla^{[\nu} E^{\rho]} \nabla_{[\rho} E_{\sigma]} \nabla^{[\sigma} E^{\mu]} \right)
\,.
\end{equation}
These couplings can be described in a way completely analogous to the $R^4$ term, through a
homogeneous real function corresponding to \eqref{eq:R4-fun}, as
\begin{equation}\label{eq:H8-fun}
 \cH_{H^{4}}= \frac{3}{16}\,\mathrm{e}^{2\,\cK(Y ,\bar{Y})}\, A_{\sf t} \bar{A}_{\sf t} \,.
\end{equation}
The final possible combination at the eight derivative level for NS fields is the $R^2 H^4$
coupling, which in 4D is characterised by the term
\begin{equation}
 \label{eq:R2H4-red}
 t_8t_8 R^4 \rightarrow 
-96\,(\nabla_{[\mu} E_{\nu]_-} \nabla^{[\mu} E^{\nu]})^2 
 R^-_{\kappa\lambda}{}^{\rho\sigma}  R^-{}^{\kappa\lambda}{}_{\rho\sigma}
\,.
\end{equation}
These couplings can be described by the obvious mixed combination of the two
functions \eqref{eq:R4-fun} and \eqref{eq:H8-fun} above, as
\begin{equation}\label{eq:R2H4-fun}
 \cH_{R^2H^4} = \frac{3}{16}\,\mathrm{e}^{2\,\cK(Y ,\bar{Y})}\, A_{\sf w} \bar{A}_{\sf t} + \text{c.c.}\,.
\end{equation}
The last function can be straightforwardly added to the functions above, to define a total
function of the vector multiplet moduli and Weyl and tensor multiplet backgrounds, defined as
\begin{equation}\label{eq:NS8-fun}
 \cH_{NS}^{8}= \frac{3}{16}\,\mathrm{e}^{2\,\cK(Y ,\bar{Y})}\, |A_{\sf w} + A_{\sf t}|^2\,,
\end{equation}
describing the eight derivative couplings of NS sector fields.
At this point it is worth pausing, to note that the form of these equations exhibits
a correspondence between the graviton and the B-field, since the complete eight
derivative action for the gravity and tensor multiplet is controlled by the
combination $A_{\sf w} + A_{\sf t}$. This will appear in several instances below, at
all orders of derivatives, and reflects the structure of the 10D Lagrangian, which
is controlled by the combination $R(\Omega_+)$.

\subsection*{Couplings involving vector multiplets}

We now turn to some of the eight derivative terms involving derivatives on vector
multiplet fields. We start with mixed terms between NS and RR fields, namely the ones
where the order of derivatives is balanced between the two sectors. Indeed, it is
straightforward to obtain the function characterising the $R^2 F^4$ coupling, which
in 4D is described by the cubic invariant in appendix \ref{app:kinetic}, where one
considers one of the chiral multiplets to be the Weyl multiplet. The $R^2 F^4$ 
coupling is then characterised by the terms
\begin{align}\label{eq:R2F4-4dterms}
&\,\cH^{(8)}_{A_{\sf w} A\, \bar B}
R^{-}_{\mu\nu\rho\lambda} R^{-}{}^{\mu\nu\rho\lambda} \,\left(
\nabla F^{-\, A} \nabla F^{+\,\bar{B}} + \Box X^{A} \Box X^{\bar{B}}
\right)
\nonumber\\
&\,+\cH^{(8)}_{A_{\sf w} A B\, \bar C \bar D}
R^{-}_{\mu\nu\rho\lambda} R^{-}{}^{\mu\nu\rho\lambda} \,
 F^{-\,A} F^{-\,B}  F^{+\,\bar{C}} F^{+\,\bar{D}}
+ \dots\,.
\end{align}
In this case, only the Ricci-like terms
contribute to the reduction altogether, so that we obtain for the relevant coupling
\begin{equation}
 \cH^{(8)}_{A_{\sf w} I\, J} = -576\,\int_X \omega_I{}^{m n}  \omega_J{}_{m n} \equiv -576\, G_{IJ}\,,
\end{equation}
i.e. proportional to the lowest order K\"ahler metric $G_{IJ}$.

This result determines the coupling of the vector multiplet scalars and the corresponding vector fields, but
we still need to fix the couplings to the Type IIA RR gauge field, labeled by $0$ in four dimensions. These
can be derived by the observation that all field strengths can be introduced by lifting the three-form
$H_{\mu_1\mu_2\mu_3}$ to the eleven dimensional four-form field strength $G_{\mu_1\mu_2\mu_3\mu_4}$ and
reducing back on a circle, keeping all components. The result of the reduction of the four-form to 4D
gauge fields, $F^I_{\mu\nu}$, naturally leads to the combination $F^I_{\mu\nu} + u^I\, F^0_{\mu\nu}$, which
should replace the field strengths in the couplings shown above, so that the full coupling becomes
\begin{equation}\label{eq:gr-ph-exten}
\cH^{(8)}_{A_{\sf w} I\, J} \rightarrow \cH^{(8)}_{A_{\sf w} A\, \bar B}
=-576\,
\begin{pmatrix}
     G_{IJ} &  G_{IJ} u^J\\
  u^I G_{IJ} &    G_{IJ} u^I u^J
\end{pmatrix}
\,.
\end{equation}

Combined with the fact that the relevant function depends on the Weyl multiplet background only
linearly, as implied by \eqref{eq:R2F4-4dterms}, one can now integrate to obtain
\begin{equation}\label{eq:R2F4-fun}
 \cH_{R^2 F^4}= 9\,A_{\sf w} \, \mathrm{e}^{2\,\cK(Y ,\bar{Y})} \,.
\end{equation}
This form is in line with the observation that the $R^2 F^4$ coupling can
be roughly seen as the product of the chiral $R^2$ term with the real $F^4$
term. Note that, unlike for the lowest order K\"ahler potential, the
$0I$ and $00$-components of the second derivative $\cH^{(8)}_{A_{\sf w} A\, \bar B}$
in \eqref{eq:R2F4-4dterms} are physical in this case, since they describe the
couplings of the RR one-form gauge field. In fact, the coupling
$\cH^{(8)}_{A_{\sf w} A\, \bar B}$ is proportional to the real part of the
period matrix, which describes the the theta angles in the two derivative theory.

The natural extension of \eqref{eq:R2F4-fun} to a function where $A_{\sf w}$ is
replaced by $A_{\sf t}$ and thus describes couplings of the type
$(\nabla E)^2 (\nabla F)^2$ is straightforward. However, in the compactification we
consider all such terms cancel identically, in a nontrivial way. Similarly, there
are no parity odd terms of this type either, so that this particular coupling
seems to be absent in four dimensions.

The same conclusion seems to hold for terms of the type $(\nabla H)^2 H^2 F^2$,
which would in principle be characteristic of the $H^6 F^2$ term in \eqref{eq:R4-red},
even though this coupling is not known in four dimensions. Terms of this order
in fields do not appear in the odd sector as well.

Finally, we consider the purely vector multiplet eight derivative couplings, corresponding to an $F^8$ term.
This can be obtained by a trivial dimensional reduction, leading to the four
dimensional coupling
\begin{equation}\label{eq:f8-coup}
  t_8t_8 R(\Omega_+)^4 \rightarrow 
72\, \big(\int_X \omega_{I}{}^{m n}  \omega_J{}_{m n} \omega_K{}^{pq}  \omega_{L}{}_{pq} \big)\,
\partial^{\mu\nu}u^{(I} \partial_{\mu\nu}u^J \partial^{\rho\sigma}u^K \partial_{\rho\sigma}u^{L)}\,,
\end{equation}
which can be described by the second quartic invariant in \eqref{eq:chiral-n4}.
Since the coupling above is given purely in terms of the product of the $(1,1)$ forms, $\omega_{I} \cdot\omega_J$,
the relevant real function is related to the K\"ahler potential and is given by 
\begin{equation}\label{eq:F8-fun}
 \cH_{F^8}=  6\,\mathrm{e}^{4\,\cK(Y ,\bar{Y})} \,.
\end{equation}
This function is consistent with \eqref{eq:f8-coup} for the $I$, $J$, indices and naturally
extends to the $0$-th gauge field in four dimensions as seen above, but we have not checked
those couplings explicitly.

\section{Six derivative couplings}
\label{sec:sixder}

At the six derivative level, we need to saturate two of the derivatives in the
internal directions, so that exactly one Riemann tensor will appear in the
relevant integrals on the Calabi-Yau manifold. This requirement turns out to
be quite restrictive, since all traces of the Calabi-Yau curvature vanish. It
follows that the internal integrals must also involve harmonic forms on which
the indices of the Riemann tensor are contracted. Given that we do not consider
any complex structure deformations, this observation directly
implies that no invariants involving only NS-NS fields, such as $R^2 H^2$ or
$H^6$, can arise in four dimensions\footnote{Note that these will become
nontrivial if more hyper/tensor multiplets are included in the reduction.}.

However, mixed couplings involving fields from both the NS-NS and the R-R sector
are nontrivial and a priori include three types of couplings, namely $R^2 F^2$,
$(\nabla E)^2 F^2$ and $H^2 F^4$. The latter has not been described in the context
of $\cN\!=\!2$ supergravity, while the former two can be constructed using the techniques
in \cite{deWit:2010za}. In addition, a purely vector multiplet coupling including six
derivatives on the component fields, i.e. an $F^6$ invariant arises.

In particular, the $R^2 F^2$ term was already constructed explicitly in \cite{deWit:2010za},
and is governed by a function, $\cH(X, A_{\sf w}; \bar{X})$, that is linear in the Weyl
multiplet, while the vector multiplet scalars appear through a holomorphic function of
degree $-2$ and an anti-holomorphic function of degree $0$. The relevant $R^2 F^2$ coupling is
\begin{equation}
\cH_{A_{\sf w}\, \bar A \bar B}
R^{-}_{\mu\nu\rho\lambda} R^{-}{}^{\mu\nu\rho\lambda} \,
F^{+\,\bar{A}}{}_{\kappa \sigma} F^{+\,\bar{B}}{}^{\kappa \sigma} + \dots\,,
\end{equation}
where the part of the coupling coming from the R-R fields, as derived
from the reduction is
\begin{equation}\label{eq:RIJ-def}
\cH_{A_{\sf w}\, \bar I \bar J}=\,
-48\,\int_{X} R_{mnpq}\,\omega_{I}{}^{mn}\omega_{J}{}^{pq} 
\equiv -48\, \cR_{IJ}\,,
\end{equation}
where in the second equality we defined the tensor $\cR_{IJ}$ for later convenience.
This tensor clearly describes a non-topological coupling, since it depends on the curvature of
the Calabi-Yau manifold explicitly. In fact, the definition \eqref{eq:RIJ-def} is invertible,
as one can reconstruct the Riemann tensor $R_{mnpq}$ from $\cR_{IJ}$ by contracting with
the harmonic two-forms. We record the following properties of $\cR_{IJ}$, which will be useful
in the discussion below,
\begin{equation}
 \cR_{IJ}=\cR_{JI}\,, \qquad t^I\cR_{IJ}=0\,, \qquad G^{IJ}\cR_{IJ}=0\,,
\end{equation}
where $t^I$ and $G^{IJ}$ are the K\"ahler moduli and $G^{IJ}$ is the inverse of the K\"ahler
metric.

In order to extend \eqref{eq:RIJ-def} to include the $0$-th gauge field, we follow the same
procedure as in \eqref{eq:gr-ph-exten}, to obtain the additional couplings
\begin{equation}
\cH^{(6)}_{\bar{A}_{\sf w}\, A B} =
\begin{pmatrix}
     \cR_{IJ} &  \cR_{IJ} u^J\\
  u^I \cR_{IJ} &    \cR_{IJ} u^I u^J
\end{pmatrix}
\rightarrow
\begin{pmatrix}
     \cR_{IJ} &  \cR_{IJ} \R( z^J)\\
  \R(z^I) \cR_{IJ} &    \cR_{IJ} \R(z^I) \R(z^J)
\end{pmatrix}
\,.
\end{equation}
We then obtain for the function describing the $R^2 F^2$ invariant
\begin{equation}\label{eq:R2F2-fun}
 \cH_{R^2F^2} =\,  
-\frac{3\mathrm{i}}{8}\,\frac{\bar{A}_{\sf w}}{(\bar{Y}^0)^2}  \hat{\cR}_{IJ}
\left( \frac{Y^I}{Y^0} - \frac{\bar{Y}^I}{\bar{Y}^0}\right)\,\left( \frac{Y^J}{Y^0} + \frac{\bar{Y}^J}{\bar{Y}^0}\right)\,,
\end{equation}
where $\hat{\cR}_{IJ}(Y, \bar{Y})=\cR_{IJ}(t)$ is viewed as a function of the $t^I=\I(\tfrac{Y^I}{Y^0})$,
as obtained in the standard special coordinates. Note that \eqref{eq:R2F2-fun} is manifestly homogeneous
in the holomorphic scalars $Y^A$, but non-homogeneous in the anti-holomorphic scalars $\bar{Y}^A$, as expected.

It is straightforward to obtain a term of the type $(\nabla E)^2 F^2$ by simply replacing
$A_{\sf w} \rightarrow A_{\sf t}$ in \eqref{eq:R2F2-fun}, in line with previous observations.
It turns out that this invariant is also generated by the reduction, as
\begin{equation}
\cH_{A_{\sf t}\, \bar A \bar B}
\nabla^{[a} E^{b]^{-}} \nabla_{[a} E_{b]^{-}} \,
F^{+\,\bar{A}}{}_{\kappa \sigma} F^{+\,\bar{B}}{}^{\kappa \sigma} + \dots\,,
\end{equation}
where the two couplings $\cH^{(6)}_{A_{\sf t}\, \bar I \bar J} = \cH^{(6)}_{A_{\sf w}\, \bar I \bar J}$,
are equal. By the same argument as above, the function \eqref{eq:R2F2-fun} can be extended
to include the tensor multiplet coupling as
\begin{equation}\label{eq:RP2F2-fun}
 \cH^{(6)}(\bar{A}_{\sf w}, \bar{A}_{\sf t}, Y, \bar{Y} ) =\,
-\mathrm{i}\,\frac{\bar{A}_{\sf w} + \bar{A}_{\sf t}}{(\bar{Y}^0)^2}  \hat{\cR}_{IJ}
\left( \frac{Y^I}{Y^0} - \frac{\bar{Y}^I}{\bar{Y}^0}\right)\,\left( \frac{Y^J}{Y^0} + \frac{\bar{Y}^J}{\bar{Y}^0}\right)\,,
\end{equation}
which describes the first row in the six-derivative part of table \ref{tbl:sum}.

We now turn to the $F^6$ term, which is computationally more challenging than the
couplings described above. This is due to the fact that there are no terms cubic in the
two-form field strength $H$ in ten dimensions, so that $(\nabla F)^3$ terms do not arise
in four dimensions. This is consistent with the fact that similar terms cancel in the
$F^6$ coupling that follows from the cubic invariant described in appendix \ref{app:kinetic}. 
One therefore is forced to consider terms of the type $(\nabla F)^2 F^2$, which are
quartic in the $(1,1)$ forms $\omega_I$, from the point of view of the Calabi-Yau reduction.

The result is a coupling containing all possible combinations of an internal Riemann tensor
and four $\omega_I$, as in
\begin{equation}
 \omega_I{}^{mn} \omega_J{}_{mn} R^{pqrs}\omega_K{}_{pq}\omega_L{}_{rs}\,,
\quad
 \omega_I{}^{mn} \omega_J{}_{np} R^{pqrs}\omega_K{}_{qm}\omega_L{}_{rs}\,,
\quad
\dots\,,
\end{equation}
which in principle determine the function controlling the $F^6$ coupling. However, we also
find nontrivial odd terms for the scalars resulting from \eqref{eq:x8}, in contrast to the
known coupling in section \ref{app:kinetic}. These terms include 
\begin{eqnarray}\label{eq:F6-odd}
\mathcal{L}_{\text{odd}} &\sim& Y_{IJKMN}\,d u^I \wedge d u^J \wedge d u^K  \wedge (\partial^\mu u^M d\, \partial_\mu u^N )\,, 
 \nn\\
 Y_{IJKMN}\,&=& \int \omega_I\wedge\omega^m{}_M\wedge\omega^n{}_N \wedge 
 \left( 2\,R_{np} \omega^{pq}{}_J \omega_{qm}{}_K  + \omega_{np}{}_J R^{pq} \omega_{qm}{}_K \right) + \dots \,,
\end{eqnarray}
where the dots stand for terms containing the same objects in double traces rather than s single one.
We observe that a term completely antisymmetric in three indices $I\,, J\,, K$ arises and conclude
that the known coupling is not sufficient to describe these terms. We leave it to future
work to determine the possible new coupling(s) that can complete the structure.

Finally, it is worth discussing in brief the invariant of the type $H^2 F^4$, which
is not known explicitly in supergravity. Such terms do appear and seem to be controlled
by the same tensor $\cR_{IJ}$ in \eqref{eq:RIJ-def} above, since we find the characteristic
couplings $\cR_{IJ} E^2 \nabla F^I \nabla F^I$ for all possible contractions of indices
between the vector and tensor multiplet field strengths. Similarly, we find the parity odd
terms
\begin{gather}
    W_{IJKL}\,H\wedge (\partial^\mu u^I d \partial_\mu u^J ) (\partial^\nu u^K \partial_\nu u^L )\,,
 \nn\\
 W_{IJKL} = \int_X \! \omega^m{}_I\wedge\omega^n{}_J \wedge  
  \omega_m{}_K\wedge\omega^p{}_L \wedge R_{np} + \dots\,,
\label{eq:H2F4-odd}
\end{gather}
where in the last integral we used similar conventions as in \eqref{eq:F6-odd} above.
Indeed, the two integrals appear to be closely related, so that the two couplings may
have a similar origin in terms of superspace invariants.

\section{Four derivative couplings}
\label{sec:fourder}

In order to obtain four derivative couplings in four dimensions from the 10D
$R^4$ invariant, one needs to consider terms that include exactly two Riemann
tensors in the internal directions. It follows that the integrals controlling
the 4D couplings are quadratic in the Calabi-Yau curvature, in the same way as
the six derivative couplings of the previous section are controlled by the
Calabi-Yau Riemann tensor through \eqref{eq:RIJ-def} above.

At this level in derivatives, four structures can appear, namely $R^2$, $F^4$,
$H^2 F^2$, and $H^4$. Given our assumption of no hyper/tensor multiplets other than
the universal tensor multiplet, all of these structures will be described by
functions involving vector multiplet scalars, but only the latter two involve the
tensor multiplet explicitly. All couplings except the $H^2 F^2$ terms can be
described straightforwardly in $\cN\!=\!2$ supergravity, and we now discuss each in turn.

\subsubsection*{The $R^2$ term}
The $R^2$ term has been known for quite some time \cite{Bergshoeff:1980is, Antoniadis:1997eg}, and arises from
terms that can be completely factorised in internal and external indices, as
\begin{align}
\label{eq:gr-te}
(t_8t_8 -\tfrac18\epsilon_{10}\epsilon_{10})R(\Omega_+)^4 \rightarrow 
\,& \alpha_I t^I\,  \left( R_{\mu \nu \rho \lambda} (\omp)  R^{\mu \nu \rho \lambda} (\omp)  + \epsilon_{abcd} R^{ab}_2(\omp) \wedge R^{cd}_2(\omp) \right) 
\nn\\
B \wedge t_8 \left[ R^4 (\Omega_+) + R^4(\Omega_-) \right] \rightarrow 
\,&\alpha_I u^I \,\big( \tr R(\Omega_+)\wedge R(\Omega_+) + \tr R(\omm)\wedge R(\omm) \big)
\end{align}
where we used \eqref{eq:zero-id} and
\begin{equation}
\alpha_I = \int_X \omega_I \wedge \tr R^2\,,
\end{equation}
are the the second Chern classes of the Calabi-Yau four-cycles. Note that this is a topological
quantity, unlike the objects controlling higher derivative couplings described above, as e.g. in \eqref{eq:RIJ-def}.

The supergravity description requires to allow the lowest order prepotential to
depend on the Weyl multiplet through $A_{\sf w}$ \cite{deWit:1996ix}, so that the
explicit prepotential arising from \eqref{eq:gr-te} is given by
\begin{equation}\label{eq:pert-prepot}
 F = -\frac16\, \frac{C_{IJK} Y^I Y^J Y^K}{Y^0} + \frac{1}{24\cdot 64}\,\frac{\alpha_I Y^I}{Y^0}\, A_{\sf w}\,,
\end{equation}
where we remind the reader that the physical moduli are given by $z^I=\frac{Y^I}{Y^0}$ in terms of the
scalars $Y^A$ above.

\subsubsection*{The $H^4$ term}
Turning to the tensor multiplet sector, an explicit computation using \eqref{eq:two-id} leads to
the following terms  in four dimensions 
\begin{align}
(t_8 t_8 - \frac18\,\varepsilon_{10}\varepsilon_{10})R^4 =&\,
48\, R^{m n p q} R_{m n p q}\,\big(
2\, {\partial}^{[\mu} E^{\nu]}\, {\partial}_{[\mu} E_{\nu]}
+\frac34\, (E^{\mu}E_{\mu})^2
\big)
\,,
\end{align}
where, in complete analogy with the $R^2$ terms, only factorised traces contribute.
It follows that the four dimensional Lagrangian contains the four derivative tensor
multiplet invariant arising from \eqref{eq:CT}, controlled by exactly the same prepotential
in \eqref{eq:pert-prepot}, upon extending the term containing the Weyl background to include
the tensor background, as
\begin{equation}\label{eq:pert-prepot-H}
 F = -\frac16\, \frac{C_{IJK} Y^I Y^J Y^K}{Y^0} + \frac{1}{24\cdot 64}\,\frac{\alpha_I Y^I}{Y^0}\,  (A_{\sf w} + 8\,A_{\sf t})\,,
\end{equation}
with $A_{\sf t}$ as in \eqref{eq:A-tens-sq}. This function describes the couplings in the
first line of Table \ref{tbl:sum}.

Once again we observe the close relation between the $R^2$ and tensor multiplet couplings,
which are characterised by exactly the same functional form in terms of the corresponding
chiral backgrounds. This structure arises despite the fact that in $\cN\!=\!2$ supergravity
in four dimensions the tensor $H$ and gravity are not in the same multiplet anymore, so that,
a priori, more flexibility, parametrized by two functions is allowed. However, we find that
the Calabi-Yau reduction leads to a single function of vector multiplet scalars for both couplings.

\subsubsection*{The $F^4$ term}

In the purely RR sector, an invariant quartic in derivatives on the vector multiplet components
exists in 4D, which is characterised by an $(\nabla F)^2$ coupling. As above, we analyse the
terms arising from the odd term under pair exchange in $R(\Omega_+)$ in order to obtain these
explicitly. Using \eqref{eq:two-id}, the terms coming from non-Ricci combinations cancel, and
the remaining ones are Laplacians of four dimensional fields. Explicitly, we obtain that the
total $(\nabla F)^2$ term reads
\begin{align}\label{eq:two-index}
(t_8 t_8 - \frac18\,\epsilon_{10}\epsilon_{10})R^4
\rightarrow&\,
 3\,X_{IJ} \nabla^2 u^I  \nabla^2 u^J\,,
\end{align}
where $X_{IJ}$ is the tensor
\begin{eqnarray}\label{eq:X-def}
X_{IJ} &=& \int_X \epsilon_{m n m_1 \ldots m_4} \epsilon_{pq n_1 \ldots n_4}
R^{m_1m_2n_1n_2}R^{m_3 m_4 n_3n_4} \,\,  \omega_I\,^{mn} \, \omega_J\,^{pq}\,,
\end{eqnarray}
and is explicitly given by
\begin{align}
 X_{IJ} =\,  8\,
\, \big[ &  ({R}_{m n p q})^2 \omega_I{}^{rs} \omega_J{}_{rs}\,
- 8\, {R}^{m n p q} {R}_{m n p}\,^{r} \omega_I{}_{q}\,^{s} \omega_J{}_{rs}
+ 4\, {R}^{m n p q} {R}_{m n}\,^{r s} \omega_I{}_{p r}\,  \omega_J{}_{q s}\,
\nonumber\\
&\,
 + 2\, {R}^{m n p q} {R}_{m n}\,^{r s} \omega_I{}_{p q}\,  \omega_J{}_{r s}\,
- 8\, {R}^{m n p q} {R}_{m}\,^{r}\,_{p}\,^{s} \omega_I{}_{n r}\,  \omega_J{}_{q s}
\big]
\end{align}
The gauge field partner of these scalar couplings is obtained by lifting to eleven
dimensions the original expression and reducing back on a circle times a Calabi-Yau.
It then follows that the result for gauge fields takes the
form
\begin{align}\label{eq:two-index-vec}
(t_8 t_8 - \frac18\,\epsilon_{10}\epsilon_{10})R^4
\rightarrow&\,
 3\,X_{IJ} \nabla^a F_{ac}^I  \nabla^b F_b{}^c{}^J\,.
\end{align}
Comparing \eqref{eq:two-index} and \eqref{eq:two-index-vec} to the known $F^4$
term in supergravity, given in \eqref{eq:real-susp-action}, we find that the
interactions of the vector multiplets arising from expansion along the second
cohomology are governed by the tensor $X_{IJ}$ above.

We now turn to the three-index structure, and compute the parity odd term quadratic in 4D field
strengths. Following the same lifting and reducing procedure, we
find that the only even-odd term quadratic in 3-from field strengths is
\begin{equation}
 t_8{}_{\mu_1\dots \mu_8} B \wedge \nabla G^{\mu_1 \mu_2 \mu_9} \wedge \nabla G^{\mu_3 \mu_4}{}_{\mu_9} \wedge R^{\mu_5 \mu_6} \wedge R^{\mu_7 \mu_8} \,,
\end{equation}
which upon reduction to 4D gives rise to a term of the type
\begin{equation}
 6\,u^K\,H_{IJ,K}\,\epsilon^{\mu\nu\rho\lambda} \nabla_{\mu} F^I{}_{\nu}{}^{\kappa} \nabla_{\rho} F^I{}_{\lambda \kappa}
\simeq
 -3\,H_{IJ,K}\,\epsilon^{\mu\nu\rho\lambda} \nabla_{\mu}u^K\, F^I{}_{\nu\rho} \nabla_{\kappa} F^I{}_{\lambda \kappa}
+\dots\,,
\end{equation}
where in the second step we partially integrated and the dots denote terms involving the derivative
of the coupling $H_{IJ,K}$. The explicit expression for this three index coupling follows from the
relation
\begin{align}
 u^K\,H_{IJ,K}=  -16\,
\, \int_{X} B\wedge & \big[ {R}^{m n}\wedge {R}_{m n} \omega_I{}^{rs} \omega_J{}_{rs}\,
- 8\, {R}^{p q}\wedge  {R}_{p}\,^{r} \omega_I{}_{q}\,^{s} \omega_J{}_{rs}
\nonumber\\
&\,
- 4\, {R}^{p q} \wedge {R}^{r s} \omega_I{}_{p r}\,  \omega_J{}_{q s}\,
 + 2\, {R}^{p q} \wedge {R}\,^{r s} \omega_I{}_{p q}\,  \omega_J{}_{r s}\, \big]\,,
\end{align}
where we note the important identities
\begin{align}
 t^K\,H_{IJ,K}
= - X_{IJ}\,,
\qquad
t^I X_{IJ} = 32\,\alpha_I\,.
\end{align}

The remaining interactions with the ten dimensional RR gauge field, $F^0$, described
by $\cH_{0I}$ and $\cH_{00}$, are obtained by viewing $F^0$ as a Kaluza-Klein gauge
field, coming from the reduction from 11D. As these must necessarily be quadratic
in the Kaluza-Klein gauge fields, only the factorised term in the 10D invariant
contributes. It then follows that, as far as terms quadratic in Riemann tensors are
concerned, the lifting and reducing procedure is identical to the 4D/5D connection
studied in \cite{Banerjee:2011ts}. Therefore, we can simply add the couplings $\cH_{0I}$
and $\cH_{00}$ found in that work, given by
\begin{align}
  \label{eq:4D-5D-res}
  \mathcal{H}_{0\bar{I}}\big|_{\text{\tiny KK}}
  =&\, -12\, \mathrm{i} \alpha_I = -\frac38\,\mathrm{i}\, X_{IJ}t^J \,,
  \nonumber\\
  \mathcal{H}_{0\bar{0}}\big|_{\text{\tiny KK}}
  =&\,24\, \alpha_I t^I = \frac34\, X_{IJ} t^I t^J \,,
\end{align}
to the ones in \eqref{eq:two-index} and \eqref{eq:two-index-vec} above.

After adding the extra contribution in \eqref{eq:4D-5D-res} to \eqref{eq:two-index}, and
performing the by now standard shift in \eqref{eq:gr-ph-exten} to account for the
axionic coupling to the $0$-th gauge field strength, we obtain the final form of the
coupling $\cH_{AB}$, as
\begin{equation}\label{eq:F4-metric}
 \cH_{A\bar{B}} =
\begin{pmatrix}
      X_{IJ} & - X_{IJ} z^J\\
 -\bar{z}^I X_{IJ} &   X_{IJ} z^I \bar{z}^J
\end{pmatrix}
\rightarrow
|Y^0|^{-4}
\begin{pmatrix}
   |Y^0|^2\, X_{IJ} & - \bar{Y}^0\, X_{IJ} Y^J\\
  - Y^0\,\bar{Y}^I  X_{IJ} &  X_{IJ} Y^I \bar{Y}^J
\end{pmatrix}\,,
\end{equation}
where in the second step we passed from the special coordinates $z^I$ to the
projective coordinates $Y^A$. Note that the coupling $X_{IJ}$ is real and
depends only on the K\"ahler moduli $t^I$, similar to the lowest order K\"ahler
potential.

The couplings \eqref{eq:F4-metric} satisfy the condition $Y^A \cH_{A\bar{B}}=0$,
so that they belong to the class of \cite{deWit:2010za}. Recently, a more general class
of $F^4$ invariants appeared in \cite{Butter:2013lta}, which allows for $Y^A \cH_{A\bar{B}}\neq 0$
and contains additional terms quadratic in the Ricci tensor. However, we find that no such extra
terms appear in the reduction of the 10D action, beyond the one in the familiar (Weyl)$^2$ term,
consistent with the properties of $\cH_{A\bar{B}}$ above.

\subsubsection*{On $H^2F^2$ terms}
\label{sec:H2F2}

Finally, we comment on the possible four derivative terms which mix tensor and vector
multiplets. Such terms have not been explicitly constructed in the literature and it
is a interesting open problem to tackle, even for rigidly supersymmetric theories. Indeed,
a construction of such an invariant is likely to lead to insight into more general mixed
terms of the type $H^{2n} F^{2m}$, where $n$ is odd, examples of which have been mentioned
above (e.g. the $H^{2} F^{4}$ term).

An explicit computation of the terms arising from reduction of the parity even terms at this
order reveals that terms involving derivatives of $H$ and $F$ do not arise. However, we do find
nontrivial terms involving field strengths only, e.g.
\begin{equation} \label{eq:H2F2-even}
 \mathcal{L} \propto X_{IJ} E^\mu E^\nu \partial_\mu u^I \partial_\nu u^J\,,
\end{equation}
and terms related to this by introducing the gauge field strengths, i.e.
$X_{IJ} E^\mu E^\nu F^I{}_{\mu}{}^{\rho} F^J{}_{\nu \rho}$, where $X_{IJ}$ is the integral
defined in \eqref{eq:X-def} above. In order to obtain this result, we used \eqref{eq:two-id}
and we note that the additional terms $\Delta J_0(\Omega_+, H)$ in \eqref{eq:ooLaghat} are
nontrivial in this case. 

In addition, the parity odd terms are also nontrivial for these couplings,
since one can easily verify that the parity odd term \eqref{eq:x8} leads to couplings of
the type
\begin{equation}\label{eq:H2F2-odd}
 Y_{IJ}\, H \wedge \partial^\mu u^{[I} d \partial_\mu u^{J]} 
\,,
\end{equation}
where $Y_{IJ}$ is the integral
\begin{equation}
 Y_{IJ}= \int \left(
   R^{mn}\wedge R_{np}\wedge \omega_I{}^p \wedge \omega_J{}_m
  -\frac1{8} \, R^{mn}\wedge R_{mn}\wedge \omega_I{}^p \wedge \omega_J{}_p \right) \,.
\end{equation}
In the last relation, the two-forms $\omega_I$ are viewed as vector valued one-forms, for
convenience. Note that the term \eqref{eq:H2F2-odd} is linear in the tensor field strength,
unlike the parity even coupling \eqref{eq:H2F2-even}. This may seem counterintuitive, but
we stress that our simplifying choice of ignoring the scalars in the tensor
multiplet may obscure the connection between tensor multiplet couplings that are expected
to be controlled by appropriate functions of these scalars. Finally, we point out that
$Y_{IJ}$ is by definition antisymmetric in its indices, which is similar to the corresponding
six derivative terms in \eqref{eq:F6-odd}-\eqref{eq:H2F4-odd} above. This type of odd terms
is somewhat unconventional in the $\cN\!=\!2$ setting and may point to a common origin of
all these unknown invariants.

One possible way to construct couplings of this type is to make use of the results of
\cite{Butter:2010jm}, on arbitrary couplings of vector and tensor multiplet superfields.
In terms of the superfields $G^2$ and $W^A$ describing the tensor and vector multiplets
respectively, one may consider an integral of the type\footnote{We thank Daniel Butter
for pointing out this possibility.}
\begin{equation}\label{eq:vec-ten}
 \int\! d^4\theta d^4\bar\theta \, \cH({ W,\bar W})\, G^2 \,,
\end{equation}
in order to describe couplings such as above, where the function $\cH$ must be such that
the couplings \eqref{eq:H2F2-even}-\eqref{eq:H2F2-odd} are reproduced. It is worth
mentioning that including kinetic multiplets in \eqref{eq:vec-ten} may lead to even
higher derivative couplings that can account for some of the unknown couplings pointed
out above, i.e. of the type $H^{2n} F^{2m}$, where $n$ is odd. The explicit realisation
of the possible Lagrangians following from the integral \eqref{eq:vec-ten} in components
would require the construction of a density formula for a general real multiplet of
$\cN\!=\!2$ supergravity and falls outside the scope of the present work.

\section{Some open questions}
\label{sec:open}

We shall conclude with a list of some open questions.

\medskip
\noindent
One immediate consequence of this work is the prediction of new four-dimensional higher-derivative $\cN\!=\!2$ invariants. It would be nice to be able to verify this prediction by explicitly constructing some of these terms, either using the structure in \eqref{eq:vec-ten}, or new techniques. It is interesting to point out that the new invariants involve terms, descending from the eleven-dimensional anomalous terms $C_3 \wedge X_8$,  which are top-form Chern-Simons-like couplings. Examples of these at the six-and four-derivative are discussed in sections \ref{sec:sixder} and \ref{sec:fourder} respectively. It would also be very interesting to verify whether the terms that we find to be vanishing but could in principle be nontrivial, such as the $H^6 F^2$ and $(\nabla H)^2 (\nabla F)^2$ terms, do exist or not.
Moreover, we stress that we have been focusing on the leading terms, matching to the invariants constructed
in \cite{deWit:2010za} and disregarding the possibility of more detailed structures that might appear. While we have
not found any inconsistencies, we cannot exclude the existence of subleading terms that are not captured here. For example,
the types of invariants recently constructed in \cite{Butter:2013lta} allow for additional couplings proportional to
the square of the Ricci tensor, rather than the Weyl tensor alone. 

\medskip
\noindent
There is a number of important omissions here. We have worked exclusively with one-loop terms, and avoided the discussion of the dilaton.  Our excuse can be that the tree-level terms neither survive the eleven-dimensional limit, nor contribute to the well studied $R^2$ terms in four dimensions. Yet they are important for understanding the corrections to the moduli spaces. In addition the dilaton is subtle and important enough to merit a discussion.

As already mentioned, we have largely ignored the complex deformations of the internal CY. It might be of some interest to extend our results to generic hyper-matter, since that would most likely turn on the couplings that we find to be vanishing.

We have concentrated only on CY compactification and hence ungauged $\cN\!=\!2$ theories. Quantum corrections to the super potential have been much studied and are of obvious interest. It would be very interesting to extend the discussion of (at least some of ) the higher derivative couplings to the gauged theories. The fact that the couplings described here have an off-shell formulation is helpful in that respect.

\medskip
\noindent
The relation of our calculation to the topological string calculations needs further elucidation. Most of our CY integrals are not topological and one may ask if there is an extension or refinement of topological strings that may capture the physical string theory couplings described here. Our calculations are exclusively one-loop, but one might hope that the structure of the terms discussed here, and the relations between different supersymmetric invariants are sufficiently restricted by supersymmetry to extend to all genus calculations.

\medskip
\noindent
The structure of the various functions describing the coupling of the gravity and tensor multiplets seem to treat the two backgrounds on the same footing, somehow reflecting the structure of the ten dimensional action built out of the torsionfull curvature tensor $R(\Omega_+)$. Given that this structure was instrumental in checking T-duality in \cite{Liu:2013dna}, it would be interesting to consider the properties of our couplings under the $c$-map, which is the lower dimensional analogous operation. Note that this would explicitly relate the vector and tensor multiplets, especially in view of the fact that the various couplings mix the two kinds of multiplets.

\medskip
\noindent
The new terms discussed here are not relevant for BPS black hole physics, at least at the attractor \cite{deWit:2010za,Dabholkar:2010uh,Gomes:2013cca}, as they
vanish by construction on fully BPS backgrounds and do not affect the entropy and charges. However, our results are relevant for non-BPS black holes
and may be related to the one-loop modifications to the entropy of such objects, as in \cite{Sahoo:2006rp}.

\section*{Acknowledgement}
We thank G. Bossard, D. Butter, B. de Wit, I. Florakis, Y. Nakayama, H. Ooguri, R. Savelli, S. Shatashvili,
A. Tomasiello and E. Witten for stimulating discussions.
The work of S.K. is supported by the European Research Council under the European Union's Seventh Framework Program (FP/2007-2013)-ERC Grant Agreement n. 307286 (XD-STRING). The work of RM is supported in part by ANR grant 12-BS05-003-01. 

\begin{appendix}

\section{Tensor structures in ten dimensions}
\label{sec_A:R4in10d}

We define the tensor, $t_8$, as having four antisymmetric pairs of indices and given
in terms of its contraction with an antisymmetric tensor $F^{\mu\nu}$ by
\begin{align}
t_8 F^4 = &\, 24\, \tr F^4  - 6\, (\tr F^2)^2\,.
\end{align}
Taking derivatives of this identity with respect to $F$ one can obtain the explicit
tensor $t_8$.
The Type IIA one-loop correction in ten dimensions contains terms quadratic in $t_8$ and
quartic in the modified curvature $R(\Omega_+)_{\mu_1\mu_2}{}^{\mu_3\mu_4}$. The latter
is antisymmetric in each pair of indices, but does not satisfy the Bianchi and pair
exchange identities. Considering a general tensor, $\cR$, with these symmetries, the
relevant expression reads
\begin{equation}
    t_8 t_8 \cR^4 = 192\, \cR_{1} + 384\, \cR_{2}
+ 24\, \cR_{3} + 12\, \cR_{4}  - 96\,( \cR_{5 a} + \cR_{5 b}) - 48\, (\cR_{6 a} + \cR_{6 b}) \,,
\end{equation}
where the $\cR_i$ are defined in \eqref{eq:R-struct} below. Similarly, we display for completeness
the full expression for the odd-odd term quartic in $\cR$ as
\begin{align}
-\frac18\,\varepsilon_{10} \varepsilon_{10} \cR^4 = &\,
192\,\tilde\cR_{1} + 24\,\tilde\cR_{3} + 12\, \tilde\cR_{4} - 192\, \tilde\cR_{5} - 384\, \tilde\cR_{6} - 384\, \tilde A_{7}
\nonumber\\
&\,
+ 4\, \cR\, \cR\,\left( \cR\, \cR\,
+ 6\,{\cR}^{\mu_1 \mu_2 \mu_3 \mu_4} {\cR}_{\mu_3 \mu_4 \mu_1 \mu_2}
- 24\, \cR\, \cR\, {\cR}^{\mu_1 \mu_2} {\cR}_{\mu_2 \mu_1}
\right)
\nonumber\\
&\,
+ 384\, \cR\,{\cR}^{\mu_1 \mu_2}\, \left(
{\cR}_{\mu_2}\,^{\mu_3}\,_{\mu_1}\,^{\mu_4} {\cR}_{\mu_4 \mu_3}
-{\cR}_{\mu_2}\,^{\mu_3 \mu_4 \mu_5} {\cR}_{\mu_4 \mu_5 \mu_1 \mu_3}
+\frac23\, {\cR}_{\mu_2}\,^{\mu_3} {\cR}_{\mu_3 \mu_1}
\right)
\nonumber\\
&\,
+ 32\, \cR\,{\cR}^{\mu_1 \mu_2 \mu_3 \mu_4} \, \left(
{\cR}_{\mu_3 \mu_4}\,^{\mu_5 \mu_6} {\cR}_{\mu_5 \mu_6 \mu_1 \mu_2}
-4\,{\cR}_{\mu_3}\,^{\mu_5}\,_{\mu_1}\,^{\mu_6} {\cR}_{\mu_4 \mu_6 \mu_2 \mu_5}
\right)
\nonumber\\
&\,
+ 96\, {\cR}^{\mu_1 \mu_2} {\cR}_{\mu_2 \mu_1}  \, \left(
2\, {\cR}^{\mu_3 \mu_4} {\cR}_{\mu_4 \mu_3}
- {\cR}^{\mu_3 \mu_4 \mu_5 \mu_6} {\cR}_{\mu_5 \mu_6 \mu_3 \mu_4} \right)
\nonumber\\
&\,
+ 768\, {\cR}^{\mu_1 \mu_2} {\cR}_{\mu_2}\,^{\mu_3}\,_{\mu_1}\,^{\mu_4}
\big(
{\cR}_{\mu_4}\,^{\mu_5 \mu_6 \mu_7} {\cR}_{\mu_6 \mu_7 \mu_3 \mu_5}
\nonumber
\\
&\,\hspace{4.3cm}
- {\cR}_{\mu_4}\,^{\mu_5}\,_{\mu_3}\,^{\mu_6} {\cR}_{\mu_6 \mu_5}
- 2\,{\cR}_{\mu_4}\,^{\mu_5} {\cR}_{\mu_5 \mu_3}
\big)
\nonumber\\
&\,
+384\, {\cR}^{\mu_1 \mu_2} {\cR}_{\mu_2}\,^{\mu_3}\,\left(
2\, {\cR}_{\mu_3}\,^{\mu_4 \mu_5 \mu_6} {\cR}_{\mu_5 \mu_6 \mu_1 \mu_4}
-{\cR}_{\mu_3}\,^{\mu_4} {\cR}_{\mu_4 \mu_1}
\right)
\nonumber\\
&\,
+ 384\, {\cR}^{\mu_1 \mu_2} {\cR}_{\mu_2}\,^{\mu_3 \mu_4 \mu_5} \big(
  {\cR}_{\mu_4 \mu_5 \mu_1}\,^{\mu_6} {\cR}_{\mu_6 \mu_3}
- {\cR}_{\mu_4 \mu_5}\,^{\mu_6 \mu_7} {\cR}_{\mu_6 \mu_7 \mu_1 \mu_3}
\nonumber\\
&\,\hspace{4.2cm}
+ 2\, {\cR}_{\mu_4}\,^{\mu_6}\,_{\mu_1 \mu_3} {\cR}_{\mu_5 \mu_6}
+ 4\, {\cR}_{\mu_4}\,^{\mu_6}\,_{\mu_1}\,^{\mu_7} {\cR}_{\mu_5 \mu_7 \mu_3 \mu_6} \big)
\end{align}
where $\cR_{\mu_1\mu_2}=\cR_{\mu_1\mu_3\mu_2}{}^{\mu_3}$ is a non-symmetric tensor corresponding to
the Ricci tensor and the scalar $\cR$ is its trace.
The various non-Ricci combinations appearing in both the even-even and odd-odd structures are defined as
\begin{align}\label{eq:R-struct}
\cR_{1} =&\, \tr \cR^{\mu_1 \mu_2} \cR_{\mu_2 \mu_3} \cR^{\mu_3 \mu_4} \cR_{\mu_4 \mu_1} \,,
 \qquad
\tilde\cR_{1} =  \tr {\cR}_{\mu_1 \mu_2} \tilde\cR^{\mu_2 \mu_3} {\cR}_{\mu_3 \mu_4} \tilde\cR^{\mu_4 \mu_1} ,
\nonumber\\
\cR_{2} = &\, \tr \cR^{\mu_1 \mu_2} \cR_{\mu_2 \mu_3} \cR_{\mu_1 \mu_4} \cR^{\mu_4 \mu_3} ,
\nonumber\\
\cR_{3}= &\, \tr \cR_{\mu_1 \mu_2} \cR^{\mu_3 \mu_4} \tr \cR^{\mu_1 \mu_2} \cR_{\mu_3 \mu_4} ,
\qquad
\tilde\cR_{3}= \tr \cR^{\mu_1 \mu_2} \tilde{\cR}^{\mu_3 \mu_4} \tr \cR_{\mu_3 \mu_4} \tilde{\cR}_{\mu_1 \mu_2},
\nonumber\\
\cR_{4}= &\, \tr \cR^{\mu_1 \mu_2} \cR_{\mu_1 \mu_2} \tr \cR^{\mu_5 \mu_6} \cR_{\mu_5 \mu_6} ,
\qquad
\tilde\cR_{4}= \tr \cR^{\mu_1 \mu_2} \tilde{\cR}_{\mu_1 \mu_2} \cR^{\mu_5 \mu_6} \tilde{\cR}_{\mu_5 \mu_6} ,
\nonumber\\
\cR_{5a} = &\, \tr \cR_{\mu_1 \mu_2} \cR^{\mu_2 \mu_5} \tr \cR_{\mu_5 \mu_6} \cR^{\mu_6 \mu_1} ,
\qquad
\cR_{5b} = \tr \tilde{\cR}^{\mu_3 \mu_4} \tilde{\cR}_{\mu_3 \mu_5} \tr \tilde{\cR}^{\mu_5 \mu_8} \tilde{\cR}_{\mu_4 \mu_8} ,
\nonumber\\
\tilde\cR_{5} = &\, \tr \cR^{\mu_1 \mu_2} \tilde{\cR}_{\mu_1 \mu_5} \tr \cR^{\mu_5 \mu_6} \tilde{\cR}_{\mu_2 \mu_6} ,
\nonumber\\
\cR_{6a} = &\, \tr \cR^{\mu_1 \mu_2} \cR^{\mu_5 \mu_6} \tr \cR_{\mu_1 \mu_5} \cR_{\mu_2 \mu_6} ,
\qquad
\cR_{6b}= \tr \tilde{\cR}^{\mu_3 \mu_4} \tilde{\cR}_{\mu_5 \mu_6} \tr \tilde{\cR}_{\mu_3 \mu_5} \tilde{\cR}_{\mu_4 \mu_6} ,
\nonumber\\
\tilde\cR_{6} = &\, \tr \cR_{\mu_1 \mu_2} \tilde\cR^{\mu_5 \mu_6} \cR_{\mu_8 \mu_5}{}^{\mu_7 \mu_1} \cR_{\mu_7 \mu_6}{}^{\mu_8 \mu_2},
\nonumber\\
 \tilde A_{7} = &\, \cR_{\mu_1 \mu_2}{}^{\mu_3 \mu_4} \cR_{\mu_3 \mu_5}{}^{\mu_1 \mu_6}
              \cR_{\mu_4 \mu_7}{}^{\mu_5 \mu_8} \cR_{\mu_6 \mu_8}{}^{\mu_2 \mu_7} \,,
\end{align}
for any tensor $\cR_{\mu_1 \mu_2}{}^{\mu_3 \mu_4}$ that is antisymmetric in each pair of indices, but does not satisfy the Bianchi identity and we use the shorthand notation $\tilde\cR_{\mu_1 \mu_2}{}^{\mu_3 \mu_4} = \cR^{\mu_3 \mu_4}{}_{\mu_1 \mu_2}$ in order to keep expressions compact. Note that if $\cR$ is identified
with a Riemann tensor, all tilded quantities become equal to their untilded counterparts.

\section[Off-shell N=2 supergravity and chiral multiplets]{Off-shell $\cN\!=\!2$ supergravity and chiral multiplets}

In this appendix we summarise some general formulae on the $\cN\!=\!2$ Weyl multiplet in
four dimensions and the chiral multiplets in a general superconformal background.
Our conventions are as in \cite{deWit:2010za}, where the reader can find a more detailed account.

\subsection*{$\cN\!=\!2$ superconformal gravity}
\label{App:N2sugra}

The off-shell formulation of four-dimensional $\cN\!=\!2$ supergravity is based on the
Weyl multiplet of conformal supergravity, whose components are given in Table
\ref{table:weyl}. This consists of the vierbein $e_\mu{}^a$, the
gravitino fields $\psi_\mu{}^i$, the dilatational gauge field $b_\mu$, the
R-symmetry gauge fields $\mathcal{V}_{\mu i}{}^j$ (which is an
anti-hermitian, traceless matrix in the $\mathrm{SU}(2)$ indices
$i,j$) and $A_\mu$, an anti-selfdual tensor field $T_{ab}{}^{ij}$, a
scalar field $D$ and a spinor field $\chi^i$. All spinor fields are
Majorana spinors which have been decomposed into chiral
components. The three gauge fields $\omega_\mu{}^{ab}$, $f_\mu{}^a$ and $\phi_\mu{}^i$, associated
with local Lorentz transformations, conformal boosts and
S-supersymmetry, respectively, are not independent as will be
discussed later.

%
\begin{table}[t]
\renewcommand{\arraystretch}{1.5}
\begin{tabular*}{\textwidth}{@{\extracolsep{\fill}}
    |c||cccccccc|ccc||cc| }
\hline
 & &\multicolumn{9}{c}{Weyl multiplet} & &
 \multicolumn{2}{c|}{parameter} \\[1mm]  \hline \hline
 field & $e_M{}^{A}$ & $\psi_M{}^i$ & $b_M$ & $A_M$ &
 $\mathcal{V}_M{}^i{}_j$ & $T_{AB}{}^{ij} $ &
 $ \chi^i $ & $D$ & $\omega_M^{AB}$ & $f_M{}^A$ & $\phi_M{}^i$ &
 $\epsilon^i$ & $\eta^i$
 \\ [.5mm] \hline
$w$  & $-1$ & $-\tfrac12 $ & 0 &  0 & 0 & 1 & $\tfrac{3}{2}$ & 2 & 0 &
1 & $\tfrac12 $ & $ -\tfrac12 $  & $ \tfrac12  $ \\[.5mm] \hline
$c$  & $0$ & $-\tfrac12 $ & 0 &  0 & 0 & $-1$ & $-\tfrac{1}{2}$ & 0 &
0 & 0 & $-\tfrac12 $ & $ -\tfrac12 $  & $ -\tfrac12  $ \\[.5mm] \hline
 $\gamma_5$   &  & + &   &    &   &   & + &  &  &  & $-$ & $ + $  & $
 -  $ \\ \hline
\end{tabular*}
\vskip 2mm
\renewcommand{\baselinestretch}{1}
\parbox[c]{\textwidth}{\caption{\label{table:weyl}{\footnotesize
Weyl and chiral weights ($w$ and $c$) and fermion
chirality $(\gamma_5)$ of the Weyl multiplet component fields and the
supersymmetry transformation parameters.}}}
\end{table}

The infinitesimal Q, S and K transformations of the
independent fields, parametrized by spinors $\epsilon^i$ and $\eta^i$
and a vector $\Lambda_\mathrm{K}{}^A$, respectively, are as
follows,
\begin{align}
  \label{eq:weyl-multiplet}
  \delta e_\mu{}^a  =&\, \bar{\epsilon}^i \, \gamma^a \psi_{ \mu i} +
  \bar{\epsilon}_i \, \gamma^a \psi_{ \mu}{}^i \, , \nonumber\\[1mm]
  \delta \psi_{\mu}{}^{i} =&\, 2 \,\mathcal{D}_\mu \epsilon^i - \tfrac{1}{8}
  T_{ab}{}^{ij} \gamma^{ab}\gamma_\mu \epsilon_j - \gamma_\mu \eta^i
  \, \nonumber \\[1mm]
  \delta b_\mu =&\, \tfrac{1}{2} \bar{\epsilon}^i \phi_{\mu i} -
  \tfrac{3}{4} \bar{\epsilon}^i \gamma_\mu \chi_i - \tfrac{1}{2}
  \bar{\eta}^i \psi_{\mu i} + \mbox{h.c.} + \Lambda^a_K e_{\mu a} \, ,
  \nonumber \\[1mm]
  \delta A_{\mu} =&\, \tfrac{1}{2} \mathrm{i} \bar{\epsilon}^i \phi_{\mu i} +
  \tfrac{3}{4} \mathrm{i} \bar{\epsilon}^i \gamma_\mu \, \chi_i +
  \tfrac{1}{2} \mathrm{i}
  \bar{\eta}^i \psi_{\mu i} + \mbox{h.c.} \, , \nonumber\\[1mm]
  \delta \mathcal{V}_\mu{}^{i}{}_j =&\, 2\, \bar{\epsilon}_j
  \phi_\mu{}^i - 3
  \bar{\epsilon}_j \gamma_\mu \, \chi^i + 2 \bar{\eta}_j \, \psi_{\mu}{}^i
  - (\mbox{h.c. ; traceless}) \, , \nonumber \\[1mm]
  \delta T_{ab}{}^{ij} =&\, 8 \,\bar{\epsilon}^{[i} R(Q)_{ab}{}^{j]} \,
  , \nonumber \\[1mm]
  \delta \chi^i =&\, - \tfrac{1}{12} \gamma^{ab} \, \Slash{D} T_{ab}{}^{ij}
  \, \epsilon_j + \tfrac{1}{6} R(\mathcal{V})_{\mu\nu}{}^i{}_j
  \gamma^{\mu\nu} \epsilon^j -
  \tfrac{1}{3} \mathrm{i} R_{\mu\nu}(A) \gamma^{\mu\nu} \epsilon^i\nonumber\\
  &\, + D \, \epsilon^i +
  \tfrac{1}{12} \gamma_{ab} T^{ab ij} \eta_j \, , \nonumber \\[1mm]
  \delta D =&\, \bar{\epsilon}^i \,  \Slash{D} \chi_i +
  \bar{\epsilon}_i \,\Slash{D}\chi^i \, .
\end{align}
Here, $D_\mu$ denotes the full superconformally covariant derivative,
while $\mathcal{D}_\mu$ denotes a covariant derivative with respect to
Lorentz, dilatation, and chiral $\mathrm{SU}(2)\times \mathrm{U}(1)$
transformations, e.g.
\begin{equation}
  \label{eq:D-epslon}
  \mathcal{D}_{\mu} \epsilon^i = \big(\partial_\mu - \tfrac{1}{4}
    \omega_\mu{}^{cd} \, \gamma_{cd} + \tfrac1{2} \, b_\mu +
    \tfrac{1}{2}\mathrm{i} \, A_\mu  \big) \epsilon^i + \tfrac1{2} \,
  \mathcal{V}_{\mu}{}^i{}_j \, \epsilon^j  \,.
\end{equation}
Under local scale and $\mathrm{U}(1)$ transformations the various
fields and transformation parameters transform as indicated in table
\ref{table:weyl}.

The various quantities denoted by $R(\mathcal{Q})$, and appearing in
the supersymmetry variations above denote the supercovariant curvature
tensors corresponding to each generator, $\mathcal{Q}$, whose detailed definition can be
found in \cite{deWit:2010za}. Here, we only give the following
\begin{align}
  \label{eq:curvatures-4}
  R(P)_{\mu \nu}{}^a  = & \, 2 \, \partial_{[\mu} \, e_{\nu]}{}^a + 2 \,
  b_{[\mu} \, e_{\nu]}{}^a -2 \, \omega_{[\mu}{}^{ab} \, e_{\nu]b} -
  \tfrac1{2} ( \bar\psi_{[\mu}{}^i \gamma^a \psi_{\nu]i} +
  \mbox{h.c.} ) \, , \nonumber\\[.2ex]
  R(Q)_{\mu \nu}{}^i = & \, 2 \, \mathcal{D}_{[\mu} \psi_{\nu]}{}^i -
  \gamma_{[\mu}   \phi_{\nu]}{}^i - \tfrac{1}{8} \, T^{abij} \,
  \gamma_{ab} \, \gamma_{[\mu} \psi_{\nu]j} \, , \nonumber\\[.2ex]
  R(M)_{\mu \nu}{}^{ab} = & \,
  \, 2 \,\partial_{[\mu} \omega_{\nu]}{}^{ab} - 2\, \omega_{[\mu}{}^{ac}
  \omega_{\nu]c}{}^b
  - 4 f_{[\mu}{}^{[a} e_{\nu]}{}^{b]}
  + \tfrac12 (\bar{\psi}_{[\mu}{}^i \, \gamma^{ab} \,
  \phi_{\nu]i} + \mbox{h.c.} ) \nonumber\\
& \, + ( \tfrac14 \bar{\psi}_{\mu}{}^i   \,
  \psi_{\nu}{}^j  \, T^{ab}{}_{ij}
  - \tfrac{3}{4} \bar{\psi}_{[\mu}{}^i \, \gamma_{\nu]} \, \gamma^{ab}
  \chi_i
  - \bar{\psi}_{[\mu}{}^i \, \gamma_{\nu]} \,R(Q)^{ab}{}_i
  + \mbox{h.c.} ) \,,
\end{align}
which are necessary to introduce the conventional constraints
\begin{align}
  \label{eq:conv-constraints}
  &R(P)_{\mu \nu}{}^a =  0 \, , \nonumber \\[1mm]
  &\gamma^\mu R(Q)_{\mu \nu}{}^i + \tfrac32 \gamma_{\nu}
  \chi^i = 0 \, , \nonumber\\[1mm]
  &
  e^{\nu}{}_b \,R(M)_{\mu \nu a}{}^b - \mathrm{i} \tilde{R}(A)_{\mu a} +
  \tfrac1{8} T_{abij} T_\mu{}^{bij} -\tfrac{3}{2} D \,e_{\mu a} = 0
  \,,
\end{align}
defining the composite gauge fields associated with local Lorentz transformations,
S-su\-per\-sym\-me\-try and special conformal boosts, $\omega_{M}{}^{AB}$,
$\phi_M{}^i$ and $f_{M}{}^A$, respectively.

\subsection*{Chiral multiplets}
\label{App:4D-chiral-multiplets}
Chiral multiplets are the basic building blocks of all supersymmetric invariants in
this paper. We therefore give a concise overview of their most basic properties, to
be used in the various constructions. 

Chiral multiplets are complex, carrying a Weyl weight $w$ and a chiral
$\mathrm{U}(1)$ weight $c$, which is opposite to the Weyl weight,
i.e. $c=-w$, while anti-chiral multiplets can be obtained from chiral
ones by complex conjugation, so that anti-chiral multiplets will have
$w=c$. The components of a generic scalar chiral multiplet are a complex
scalar $A$, a Majorana doublet spinor $\Psi_i$, a complex symmetric
scalar $B_{ij}$, an anti-selfdual tensor $G_{ab}^-$, a Majorana
doublet spinor $\Lambda_i$, and a complex scalar $C$. The assignment
of their Weyl and chiral weights is shown in
table~\ref{table:chiral}.
The Q- and S-supersymmetry transformations
for a scalar chiral multiplet of weight $w$, are as
follows
\begin{align}
  \label{eq:conformal-chiral}
  \delta A =&\,\bar\epsilon^i\Psi_i\,, \nonumber\\[.2ex]
  \delta \Psi_i =&\,2\,\Slash{D} A\epsilon_i + B_{ij}\,\epsilon^j +
  \tfrac12   \gamma^{ab} G_{ab}^- \,\varepsilon_{ij} \epsilon^j + 2\,w
  A\,\eta_i\,,  \nonumber\\[.2ex]
  \delta B_{ij} =&\,2\,\bar\epsilon_{(i} \Slash{D} \Psi_{j)} -2\,
  \bar\epsilon^k \Lambda_{(i} \,\varepsilon_{j)k} + 2(1-w)\,\bar\eta_{(i}
  \Psi_{j)} \,, \nonumber\\[.2ex]
  \delta G_{ab}^- =&\,\tfrac12
  \varepsilon^{ij}\,\bar\epsilon_i\Slash{D}\gamma_{ab} \Psi_j+
  \tfrac12 \bar\epsilon^i\gamma_{ab}\Lambda_i
  -\tfrac12(1+w)\,\varepsilon^{ij} \bar\eta_i\gamma_{ab} \Psi_j \,,
  \nonumber\\[.2ex]
  \delta \Lambda_i =&\,-\tfrac12\gamma^{ab}\Slash{D}G_{ab}^-
   \epsilon_i  -\Slash{D}B_{ij}\varepsilon^{jk} \epsilon_k +
  C\varepsilon_{ij}\,\epsilon^j
  +\tfrac14\big(\Slash{D}A\,\gamma^{ab}T_{abij}
  +w\,A\,\Slash{D}\gamma^{ab} T_{abij}\big)\varepsilon^{jk}\epsilon_k
  \nonumber\\
  &\, -3\, \gamma_a\varepsilon^{jk}
  \epsilon_k\, \bar \chi_{[i} \gamma^a\Psi_{j]} -(1+w)\,B_{ij}
  \varepsilon^{jk}\,\eta_k + \tfrac12 (1-w)\,\gamma^{ab}\, G_{ab}^-
    \eta_i \,, \nonumber\\[.2ex]
    \delta C =&\,-2\,\varepsilon^{ij} \bar\epsilon_i\Slash{D}\Lambda_j
  -6\, \bar\epsilon_i\chi_j\;\varepsilon^{ik}
    \varepsilon^{jl} B_{kl}   \nonumber\\
  &\, -\tfrac14\varepsilon^{ij}\varepsilon^{kl} \big((w-1)
  \,\bar\epsilon_i \gamma^{ab} {\Slash{D}} T_{abjk}
    \Psi_l + \bar\epsilon_i\gamma^{ab}
    T_{abjk} \Slash{D} \Psi_l \big) + 2\,w \varepsilon^{ij}
    \bar\eta_i\Lambda_j \,.
\end{align}
%
\begin{table}[t]
\begin{center}
{\renewcommand{\arraystretch}{1.5}
\renewcommand{\tabcolsep}{0.2cm}\begin{tabular}{|c||cccccc| }
\hline
 & & \multicolumn{4}{c}{Chiral multiplet} & \\  \hline \hline
 field & $A$ & $\Psi_i$ & $B_{ij}$ & $G_{ab}^-$& $\Lambda_i$ & $C$ \\[.5mm] \hline
$w$  & $w$ & $w+\tfrac12$ & $w+1$ & $w+1$ & $w+\tfrac32$ &$w+2$
\\[.5mm] \hline
$c$  & $-w$ & $-w+\tfrac12$ & $-w+1$ & $-w+1$ & $-w+\tfrac32$ &$-w+2$
\\[.5mm] \hline
$\gamma_5$   & & $+$ & &   & $+$ & \\ \hline
\end{tabular}}
\vskip 2mm
\renewcommand{\baselinestretch}{1}
\parbox[c]{10.8cm}{\caption{\label{table:chiral}{\footnotesize
Weyl and chiral weights ($w$ and $c$) and fermion
chirality $(\gamma_5)$ of the chiral multiplet component fields.}}}
\end{center}
\end{table}

Any homogeneous function of chiral superfields constitutes a chiral superfield,
whose Weyl weight is determined by the degree of homogeneity of the function at
hand. Indeed, one can show that a function $G(\Phi)$ of chiral
superfields $\Phi^I$ defines a chiral superfield, whose component fields
take the following form,
\begin{align}
  \label{eq:chiral-mult-exp}
A\vert_G =&\, G \,,   \nonumber\\
 \{ \, \Psi_i\,, B_{ij}\,, G_{ab}^- \, \}\vert_G 
    = &\, G_I \,\{ \, \Psi_i{}^I\,, B_{ij}{}^I\,, G_{ab}^-{}^I \, \}
  \,,\nonumber\\
  \Lambda_{i}\vert_G =&\, G_I \,\Lambda_{i}{}^I
  -\tfrac12
  G_{IJ}\big[B_{ij}{}^I   \varepsilon^{jk} +\tfrac12\, G^{-}_{ab}{}^I\gamma^{ab}\delta_i^k \big]\,\Psi_{k}{}^J
   \,,\nonumber\\
   C\vert_G =&\, G_I\, C^I  -\tfrac14
   G_{IJ}\big[ B_{ij}{}^I B_{kl}{}^J\,
   \varepsilon^{ik} \varepsilon^{jl}
   -2\, G^{-}_{ab}{}^I G^{-abJ} \big]  \,,
\end{align}
where $G_{I}$, $G_{IJ}$ etc. are the derivatives of the function $G$ with
respect to the scalars $A^I$ and we omitted all terms nonlinear in fermions
for brevity.

%
\begin{table}[t]
\begin{center}
\renewcommand{\arraystretch}{1.5}
\begin{tabular*}{13.5cm}{@{\extracolsep{\fill}}|c||cccc||cccc| }
\hline
 & \multicolumn{4}{c||}{vector multiplet} &
 \multicolumn{4}{c|}{tensor multiplet}\\
 \hline \hline
 field & $X$ & $W_\mu$  & $\Omega_i$ & $Y^{ij}$& $L^{ij}$ &
 $B_{\mu\nu}$ & $\varphi_i$ & $G$ \\[.5mm] \hline
$w$  & $1$ & $0$ & $\tfrac32$ & $2$ & $2$& $0$ &$\tfrac52$& $3$
\\[.5mm] \hline
$c$  & $-1$ & $0$ & $-\tfrac12$ & $0$ &$0$&$0$&$-\tfrac12$ &$1$
\\[.5mm] \hline
$\gamma_5$   & && $+$  &   &&&$-$& \\ \hline
\end{tabular*}
\vskip 2mm
\renewcommand{\baselinestretch}{1}
\parbox[c]{13.5cm}{\caption{\label{table:vector}\footnotesize Weyl and chiral weights ($w$
    and $c$) and fermion chirality $(\gamma_5)$ of the vector
    multiplet and the tensor multiplet. }}
\end{center}
\end{table}

Chiral multiplets of $w=1$ are special, because they are reducible upon imposing a reality constraint.
The two cases that are relevant are the vector multiplet, which
arises upon reduction from a scalar chiral multiplet, and the Weyl multiplet, which is a reduced
anti-selfdual chiral tensor multiplet.

The constraint for a scalar chiral superfield implies that $C\vert_{\text{vector}}$ and
$\Lambda_i\vert_{\text{vector}}$ are expressed in terms of the lower components of the multiplet,
and imposes a reality constraint on $B\vert_{\text{vector}}$ and a Bianchi identity on $G^-\vert_{\text{vector}}$ \cite{Firth:1974st,deRoo:1980mm,deWit:1980tn}, as
\begin{align}
  \label{eq:vect-mult}
  A\vert_{\text{vector}}=&\,X\,,\nonumber\\
  \Psi_i\vert_{\text{vector}}=&\, \Omega_i\,,\nonumber\\
  B_{ij}\vert_{\text{vector}}=&\, Y_{ij}
  =\varepsilon_{ik}\varepsilon_{jl}Y^{kl}\,,\nonumber\\
  G_{ab}^-\vert_{\text{vector}}=&  F_{ab}^-
  -\tfrac14\,
    \bar{X}\, T_{ab}{}^{ij}\,\varepsilon_{ij}  \,,\nonumber\\
  \Lambda_i\vert_{\text{vector}}
  =&\,-\varepsilon_{ij}\Slash{D}\Omega^j\nonumber\\
  C\vert_{\text{vector}}= &\,-2\, \Box_\mathrm{c}  \bar X  -\tfrac14  G_{ab}^+\,
   T^{ab}{}_{ij} \varepsilon^{ij} 
  \,,
\end{align}
where $F_{\mu\nu}= 2 \partial_{[\mu} A_{\nu]}$ is the field strength of a gauge
field, $A_{\mu}$. The corresponding Bianchi identity on $G_{ab}$ can be written as,
\begin{align}
\label{eq:Bianchi-vector} D^b\left(G_{ab}^+-G_{ab}^- +\ft14 X
  T_{abij}\varepsilon^{ij}-\ft14 \bar{X}
  T_{ab}{}^{ij}\varepsilon_{ij}\right)
 =0 \,,
\end{align}
where in both \eqref{eq:vect-mult} and \eqref{eq:Bianchi-vector} we again omitted
terms nonlinear in fermions.
The reduced scalar chiral multiplet thus describes the
covariant fields and field strength of a {\it vector multiplet}, which
encompasses $8+8$ bosonic and fermionic
components. Table~\ref{table:vector} summarizes the Weyl and chiral
weights of the various fields belonging to the vector multiplet: a
complex scalar $X$, a Majorana doublet spinor $\Omega_i$, a vector
gauge field $A_\mu$, and a triplet of auxiliary fields $Y_{ij}$.

The Q- and S-supersymmetry transformations for the vector multiplet
take the form,
\begin{align}
  \label{eq:variations-vect-mult}
  \delta X =&\, \bar{\epsilon}^i\Omega_i \,,\nonumber\\
  \delta\Omega_i =&\, 2 \Slash{D} X\epsilon_i
     +\ft12 \varepsilon_{ij}  G_{\mu\nu}
   \gamma^{\mu\nu}\epsilon^j +Y_{ij} \epsilon^j
     +2X\eta_i\,,\nonumber\\
  \delta A_{\mu} = &\, \varepsilon^{ij} \bar{\epsilon}_i
  (\gamma_{\mu} \Omega_j+2\,\psi_{\mu j} X)
  + \varepsilon_{ij}
  \bar{\epsilon}^i (\gamma_{\mu} \Omega^{j} +2\,\psi_\mu{}^j
  \bar X)\,,\nonumber\\
\delta Y_{ij}  = &\, 2\, \bar{\epsilon}_{(i}
  \Slash{D}\Omega_{j)} + 2\, \varepsilon_{ik}
  \varepsilon_{jl}\, \bar{\epsilon}^{(k} \Slash{D}\Omega^{l)
  } \,,
\end{align}
and, for $w=1$, are in clear correspondence with the supersymmetry
transformations of generic scalar chiral multiplets given in
\eqref{eq:conformal-chiral}.

We now turn to the covariant fields of the Weyl multiplet, which can be arranged in an 
anti-selfdual tensor chiral multiplet, whose chiral superfield components take the following form,
\begin{align}
  \label{eq:W-mult}
  A_{ab}\vert_{W}   =&\,T_{ab}{}^{ij}\varepsilon_{ij}\,,\nonumber \\
  \Psi_{abi}\vert_{W} =&\, 8\, \varepsilon_{ij}R(Q)^j_{ab} \,,\nonumber\\
  B_{abij}\vert_{W}  =&\, -8 \,\varepsilon_{k(i}R({\cal V})_{ab}^-{}^k{}_{j)} \,,\nonumber\\
  \left(G^{-}_{ab}\right){}^{cd}\vert_{W}  =&\, -8
\,\hat{R}(M)_{ab}^-{}^{\!cd} \,,\nonumber\\
  \Lambda_{abi}\vert_{W} =&\, 8\left(\mathcal{R}(S)_{abi}^- + \ft34 \gamma_{ab}\Slash{D}\chi_i\right)
  \,,\nonumber\\
  C_{ab}\vert_{W} =&\,  4 D_{[a} \,D^cT_{b]c\,ij}
  \varepsilon^{ij}-\text{dual} \,.
\end{align}
Note that all quantities involved in the components above are either manifestly supercovariant
curvatures or (covariant) auxiliary fields of the Weyl multiplet. In particular,
$\mathcal{R}(S)_{abi}$ is the curvature of the S-supersymmetry gauge field, which is solved in
terms of the derivative of the gravitino curvature, $\mathcal{R}(Q)_{abi}$, due to the conventional
constraints.

All higher derivative terms involving powers of the Weyl tensor in this paper are constructed
by couplings of the scalar chiral multiplet with $w=2$ is obtained by squaring the Weyl
multiplet above. The various scalar chiral multiplet components of this multiplet are given by,
\begin{align}
  \label{eq:W-squared}
  A_{\sf w}   =&\,(T_{ab}{}^{ij}\varepsilon_{ij})^2\,,\nonumber \\[.2ex]
  \Psi_{\sf w}{}_i  =&\, 16\, \varepsilon_{ij}R(Q)^j_{ab} \,T^{klab}
  \, \varepsilon_{kl} \,,\nonumber\\[.2ex]
  B_{ij}{}_{\sf w}   =&\, -16 \,\varepsilon_{k(i}R({\cal
    V})^k{}_{j)ab} \, T^{lmab}\,\varepsilon_{lm} -64
  \,\varepsilon_{ik}\varepsilon_{jl}\,\bar R(Q)_{ab}{}^k\, R(Q)^{l\,ab}
  \,,\nonumber\\[.2ex]
  G^{-ab}_{\sf w}   =&\, -16 \,\hat{R}(M)_{cd}{}^{\!ab} \,
  T^{klcd}\,\varepsilon_{kl}  -16 \,\varepsilon_{ij}\, \bar
  R(Q)^i_{cd}  \gamma^{ab} R(Q)^{cd\,j}  \,,\nonumber\\[.2ex]
  \Lambda_i{}_{\sf w}  =&\, 32\, \varepsilon_{ij} \,\gamma^{ab} R(Q)_{cd}^j\,
  \hat{R}(M)^{cd}{}_{\!ab}
  +16\,({\cal R}(S)_{ab\,i} +3 \gamma_{[a} D_{b]}  \chi_i) \,
  T^{klab}\, \varepsilon_{kl} \nonumber\\
  &\, -64\, R({\cal V})_{ab}{}^{\!k}{}_i \,\varepsilon_{kl}\,R(Q)^{ab\,l}
  \,,\nonumber\\[.2ex]
  C_{\sf w}  =&\,  64\, \hat{R}(M)^{-cd}{}_{\!ab}\,
 \hat{R}(M)^-_{cd}{}^{\!ab}  + 32\, R({\cal V})^{-ab\,k}{}_l^{~} \,
  R({\cal V})^-_{ab}{}^{\!l}{}_k  \nonumber \\
  &\, - 32\, T^{ab\,ij} \, D_a \,D^cT_{cb\,ij} +
  128\,\bar{\mathcal{R}}(S)^{ab}{}_i  \,R(Q)_{ab}{}^i  +384 \,\bar
  R(Q)^{ab\,i} \gamma_aD_b\chi_i   \,.
\end{align}
In practice, we will only use the lowest component, $A_{\sf w}$, to construct
functions that define composite chiral multiplets, as in \eqref{eq:chiral-mult-exp},
which determines completely all instances of the higher components in the
relevant couplings. The components \eqref{eq:W-squared} can then
be substituted straightforwardly in the final expressions to obtain the
explicit couplings to the fields of the Weyl background.

\section{Tensor multiplet as a chiral background}
\label{sec:tensor}
We now turn to the tensor multiplet, which is also defined as an off-shell
multiplet in an arbitrary superconformal background. The field content of
this multiplet includes a pseudoreal triplet of scalars, $L_{ij}$, a
two-form gauge potential, $B_{\mu\nu},$ a Majorana fermion doublet,
$\varphi^i$, and an auxiliary complex scalar, $G$, with the Weyl and chiral
assignments given in \ref{table:vector}. The corresponding
supersymmetry transformation rules are as follows
\begin{equation}
  \label{eq:tensor-tr}
  \begin{split}
  \delta L_{ij} =& \,2\,\bar\epsilon_{(i}\varphi_{j)} +2
  \,\varepsilon_{ik}\varepsilon_{jl}\,
  \bar\epsilon^{(k}\varphi^{l)}  \,,\\
  \delta\varphi^{i} =& \,\Slash{D} L^{ij} \,\epsilon_j +
  \varepsilon^{ij}\,\Slash{\hat E}^I \,\epsilon_j - G \,\epsilon^i
  + 2 L^{ij}\, \eta_j \,,\\
  \delta G =& \,-2 \,  \bar\epsilon_i \Slash{D} \, \varphi^{i} \,
  - \bar\epsilon_i  ( 6 \, L^{ij} \, \chi_j + \tfrac1{4} \,
    \gamma^{ab}  T_{ab jk} \, \varphi^l \,
    \varepsilon^{ij} \varepsilon^{kl}) + 2 \, \bar{\eta}_i\varphi^{i}
    \, ,\\
  \delta B_{\mu\nu} =& \, \mathrm{i}\bar\epsilon^i\gamma_{\mu\nu}
  \varphi^{j} \,\varepsilon_{ij} - \mathrm{i}\bar\epsilon_i\gamma_{\mu\nu}
  \varphi_{j} \,\varepsilon^{ij} \,  + \,2 \mathrm{i} \, L_{ij} \,
  \varepsilon^{jk} \, \bar{\epsilon}^i \gamma_{[\mu} \psi_{\nu ]k}
  - 2 \mathrm{i}\,  L^{ij} \, \varepsilon_{jk} \, \bar{\epsilon}_i
    \gamma_{[\mu} \psi_{\nu ]}{}^k \, ,
\end{split}
\end{equation}
and we refer to \cite{deWit:2006gn} for the precise definitions of the
superconformally covariant derivatives on the various fields. The vector
$\hat E^\mu$ is the superconformal completion of the dual of the three-form
field strength, $\hat E^\mu = \tfrac{1}{2}\mathrm{i}\, e^{-1} \, \varepsilon^{\mu \nu
  \rho \sigma} \partial_\nu B_{\rho \sigma}$.

The couplings of the tensor multiplets are given in terms of composite
vector multiplets \cite{deWit:2006gn}, described by functions of a set
of tensor multiplets, labeled by $I$.
To this end, we define the first component, the scalar $X_I$ as
\begin{equation} \label{eq:X-tens}
   X_I = {\cal F}_{I,J} \, \bar G^J +
  {\cal F}_{I,JK}{}^{ij}\,\bar\varphi_i{}^J\varphi_j{}^K \,,
\end{equation}
which, by \eqref{eq:tensor-tr}, transforms according to the first
of \eqref{eq:conformal-chiral} into the remaining bosonic components of the
vector multiplet, as
\begin{eqnarray}
  \label{eq:Omega-tens}
   Y_{ij \,I} &=& -2\,{\cal F}_{I,J}\, \Big[\Box^{\rm c} L_{ij}{}^J +
   3\,D L_{ij}{}^J \Big]
   -2\,{\cal F}_{I,JKij} \,( \bar G^J\,G^K + \hat E_{\mu}{}^J\,\hat
   E^{\mu K}) \,,
  \nn\\
&&{}
  -2\, {\cal F}_{I,JK}{}^{kl}\,( D_\mu L_{ik}{}^J\,D^\mu
  L_{jl}{}^K + 2\,\varepsilon_{k(i}\, D_\mu L_{j)l}{}^J\,
  \hat E^{\mu K} )
\nn\\
\label{eq:sc-eqF}
   F_{\mu\nu\,I} &=& - 2\,{\cal F}_{I,JK}{}^{mn} \,\partial_{[\mu}
   L_{mk}{}^J \, \partial_{\nu]} L_{nl}{}^K \,\varepsilon^{kl} \nn\\
&&{}
-4\, \partial_{[\mu} \left( {\cal F}_{I,J} \,\hat E_{\nu]}{}^J
-\tfrac12\,{\cal F}_{I,J} \, \mathcal{V}_{\nu]}{}^i{}_j \,
    L_{ik}{}^J \, \varepsilon^{jk}
 \right)
 \, ,
 \nn \\
 C_I & = &\,-2\, \Box_\mathrm{c} ( {\cal F}_{I,J} \,G^J )
 -\tfrac14 \, ( F_{ab\, I}^+  -\tfrac14\,{\cal F}_{I,J} \,\bar G^J T_{ab\,ij} \varepsilon^{ij})\,
 T^{ab}{}_{ij} \varepsilon^{ij}  \,,
\end{eqnarray}
where we suppressed all fermions and the component $C_I$ is consistent with \eqref{eq:vect-mult}.
In order for this multiplet to be well defined, the first
derivative of ${\cal F}_{I,J}(L)$ with respect to
$L^{K\,ij}$, denoted by ${\cal F}_{I,J,Kij}$, must satisfy
the constraints
\begin{equation}
  \label{eq:chiral-constraints}
  {\cal F}_{I,J,Kij}= {\cal F}_{I,K,Jij}\,,\qquad
   \varepsilon^{jk}\, {\cal F}_{I,J,Kij,Lkl}(L)=0\,,
\end{equation}
while Weyl covariance requires the condition
\begin{equation}
  \label{eq:sc-tensor}
  {\cal F}_{I,JKik}\,L^{kjK} = -\tfrac1{2} \delta_i{}^j \,{\cal
  F}_{I,J}\,,
\end{equation}
which implies that the function ${\cal F}_{I,J}$ is
${\rm SU}(2)$ invariant and homogeneous of degree $-1$, so
that it has scaling weight $-2$.

The expressions for the composite chiral supermultiplet above can be
used to construct actions with higher derivative couplings.
In general, one can use \eqref{eq:X-tens}-\eqref{eq:sc-eqF} on the same
footing as any vector multiplet to obtain actions containing vector-tensor
couplings. This is beyond the scope of this paper, where we only consider
a background chiral multiplet containing four derivatives on the components
of a single tensor multiplet, similar to \cite{deWit:2006gn} but allowing
for couplings depending on vector multiplet scalars as well.

For a single tensor multiplet, the functions $\mathcal{F}_{I,J}$
in \eqref{eq:X-tens} reduce to a single function $\mathcal{F}(L)$,
while the constraints \eqref{eq:chiral-constraints}-\eqref{eq:sc-tensor}
imply the constraint,
\begin{equation}
  \label{eq:laplace-F}
  \frac{ \partial^2\mathcal{F}(L)}{\partial L^{ij}\,\partial L_{ij}} = 0
  \,.
\end{equation}
We then consider the chiral multiplet of $w=2$ defined by its first
component as the square of \eqref{eq:X-tens}, through
\begin{equation}\label{eq:A-tens-sq}
 \hat A^{\sf t} = \mathcal{F}^2 \bar G^2
+ 2\,\mathcal{F}\,\mathcal{F}^{ij}\, G \,\bar\varphi_i{}\varphi_j
= \cH\,G^2 + \cH^{ij}\, \bar G \,\bar\varphi_i{}\varphi_j
\,,
\end{equation}
where we defined the function $\mathcal{H}(L) = [\mathcal{F}(L)]^2$, and
its derivatives, as
\begin{equation}
  \label{eq:H-der}
  \mathcal{H}^{ij} = \frac{\partial\mathcal{H}}{\partial
  L_{ij}}\;,\qquad    \mathcal{H}^{ij,kl}  =
  \frac{\partial^2\mathcal{H}}{\partial L_{ij}\,\partial L_{kl}}\;.
\end{equation}
As noted in \eqref{eq:two-der-ten}, for a single tensor multiplet the
function $\mathcal{F}$ is essentially unique, so that $\mathcal{H}$ is simply
given by its square, as
\begin{align}\label{eq:F-sing-ten}
 \mathcal{H} = \frac{1}{ L_{ij} L^{ij} }\,,
\end{align}
where in the reduction we consider in the main text, the scalars $L_{ij}$ contain
the dilaton and are kept constant throughout.

The remaining components of this composite background multiplet are given by
\eqref{eq:chiral-mult-exp} for $G(A)=A^2$, as follows from \eqref{eq:A-tens-sq}.
For completeness, we display their form for a general function $\cH$, as follows
\begin{align}
  B^{\sf t}{}_{ij} =&\,
  2 \,\bar G \,
 \left( 2\,\cH\, \Big[\Box^{\rm c} L_{ij} +  3\,D L_{ij} \Big]
   -\cH_{ij} \,( |G|^2 + \hat E_{\mu}\,\hat E^{\mu})
\right. \nonumber\\
  &
\left.-\cH^{kl}\,( D_\mu L_{ik} \,D^\mu L_{jl}
  + 2\,\varepsilon_{k(i}\, D_\mu L_{j)l} \, \hat E^{\mu} ) \right)
 \,,\nonumber\\
  G^{\sf t}{}_{ab}^-{} =&\,
  - 2\,\cH^{mn} \,\bar G \, \mathcal{D}_{[a}
   L_{mk} \, \mathcal{D}_{b]} L_{nl} \,\varepsilon^{kl}
 -8\,\cH\,\bar G \, \left( \mathcal{D}_{[a}\hat E_{b]}
-\tfrac14\,  R_{ab}{}^i{}_j(\mathcal{V}) \,
    L_{ik} \, \varepsilon^{jk}  \right)  \nonumber\\
   &\,
   -4\,\bar G \cH_{mn} \, \mathcal{D}_{[a} L^{mn} \hat E_{b]}
 -\tfrac12\, \cH \, |G|^2  T_{ab}{}^{ij} \varepsilon_{ij}
 \,,\nonumber\\
\end{align}
for the lower components and
\begin{align}
  \label{eq:CT}
  C^{\sf t} = & \, \mathcal{H}(L) \Big\{
   -4 \bar G \Box_\mathrm{c} G
  - 2 \left( \Box_\mathrm{c} L_{ij} + 3\,D\,L_{ij} \right)^2
  + 16\, \mathcal{D}_{[a} E_{b]^-} \, \mathcal{D}^{[a} E^{b]^-} \nn\\
  & \qquad\quad
  - 8\,\mathcal{D}_a E_b \big( {R}^{ab}{}^{i}{}_j^-(\mathcal{V}) L_{ik}
  \varepsilon^{jk}
  - \tfrac14 [ T^{ab}{}^{ij} \varepsilon_{ij} \,G  + \mbox{h.c.}]\big)
   \nn\\[1ex]
  &\qquad\quad
  +\tfrac1{16} \left((T_{ab}{}^{ij} \varepsilon_{ij})^2 G^2
  + 2\,(T_{ab}{}_{ij} \varepsilon^{ij})^2 \bar G^2 \right)
  + 12\, (\vert G \vert^2 + E^2)\,D
   \nn\\[1ex]
  &\qquad\quad
  + {R}^{ab}{}^{m}{}_n(\mathcal{V}) L_{ml} \varepsilon^{nl}
  \big( {R}_{ab}{}^{i}{}_j^-(\mathcal{V}) L_{ik} \varepsilon^{jk}
  - \tfrac12 [T_{ab}{}^{ij} \varepsilon_{ij} \,G  + \mbox{h.c.}] \big)
     \Big\}
  \nn\\[1ex]
&\, + \mathcal{H}^{ij}(L) \Big\{
  \left( \Box_\mathrm{c} L^{kl} + 3\,D\,L^{kl} \right)
  \left(\mathcal{D}_\mu L_{ik} \mathcal{D}^{\mu} L_{jl}
  -4\, \varepsilon_{ik} E^\mu \mathcal{D}_\mu L_{jl} \right)
  \nn\\[1ex]
  & \qquad \qquad
  -2\,\Box_\mathrm{c} L_{ij} (2\,\vert G \vert^2 + E^2 )
  -4\,\bar G\, \mathcal{D}^\mu G\, \mathcal{D}_\mu L_{ij}
  \nn\\[1ex]
  & \qquad\qquad
  -4\, \big(E_b \mathcal{D}_a L_{ij}  +\tfrac12 \mathcal{D}_a L_{ik}
  \mathcal{D}_b L_{jl} \varepsilon^{kl} \big)
  \big( {R}^{ab}{}^{m}{}_n^-(\mathcal{V}) L_{mo} \varepsilon^{no}
  - \tfrac14 \,T^{ab}{}^{mn} \varepsilon_{mn} G
  \big)\nn\\[1ex]
  &\qquad\qquad
  +8 (\mathcal{D}_a L_{ik} \mathcal{D}_b L_{jl} \varepsilon^{kl}
  -2\, E_a \mathcal{D}_b L_{ij}) \mathcal{D}^{[a} E^{b]^-}
  \Big\}
  \nn\\[1ex]
& + \mathcal{H}^{ij,kl}(L) \Big\{
  - \varepsilon_{ik} \varepsilon^{pq}
   \mathcal{D}^\mu L_{mp} \mathcal{D}^\nu L^{mn}
   \mathcal{D}_\mu L_{jn} \mathcal{D}_\nu L_{ql}
  \nn\\[1ex]
  & \qquad \qquad
  - 8\, \varepsilon_{ik}
 E^b \mathcal{D}^a L_{jm}
(\mathcal{D}_a L^{mn} \mathcal{D}_b L_{nl}
+ \tfrac16 \epsilon_{abcd}\mathcal{D}^c L_{mn} \mathcal{D}^d L_{nl} )
  \nn\\[1ex]
  & \qquad\qquad
  + 2\,\mathcal{D}_\mu L_{ik} \mathcal{D}^{\mu} L_{jl} \vert G \vert^2
  -  (\vert G \vert^2 + E^2 )\,
  \left( \varepsilon_{ik} \varepsilon_{jl}(\vert G \vert^2 + E^2 )
    + 4 \varepsilon_{ik} E^\mu \mathcal{D}_\mu L_{jl} \right)
  \nn\\[1ex]
  & \qquad\qquad
  + 2 \varepsilon_{ik}\varepsilon^{mn}\mathcal{D}_\mu L_{jm} \mathcal{D}^{\mu} L_{nl}\,E^2
  + 4 \varepsilon_{ik}( \mathcal{D}_\mu L_{jm} \mathcal{D}_\nu L_{ln} \varepsilon^{mn}
  ) E^\mu E^\nu
 \Big\}  \, ,
\end{align}
for the top component.

\section{The kinetic multiplet and supersymmetric invariants}
\label{app:kinetic}

The central object in constructing the various higher derivative invariants
of the type $R^{2n} F^{2m}$ in this paper is the so called
kinetic chiral multiplet. The term `kinetic' multiplet was first used in
the context of the $N=1$ tensor calculus \cite{Ferrara:1978jt}, because
this is the chiral multiplet that enables the construction of the
kinetic terms, conventionally described by a real superspace integral,
in terms of a chiral superspace integral. In \cite{deWit:1980tn, deWit:2010za}
a corresponding kinetic multiplet, $\mathbb{T}(\bar\Phi)$, for a chiral
$w=0$ multiplet, $\Phi$, was identified for $N=2$ supersymmetry,
which now involves four rather than two covariant $\bar\theta$-derivatives.
It follows that $\mathbb{T}(\bar\Phi)$ contains up to four space-time
derivatives, so that the expression
\begin{equation}
  \label{eq:real-chiral-n2}
  \int \!\mathrm{d}^4\theta\;\mathrm{d}^4\bar\theta \;\Phi\,\bar\Phi^\prime
  \approx
  \int \!\mathrm{d}^4\theta \,\Phi\,\mathbb{T}(\bar\Phi^\prime) \,,
\end{equation}
corresponds to a four derivative coupling. Expressing the chiral
multiplets in terms of (functions of) reduced chiral multiplets,
\eqref{eq:real-chiral-n2} leads to higher-derivative couplings
of vector multiplets and/or the Weyl multiplet.

Denote the components of a $w=0$ chiral multiplet by
$(A,\Psi,B,G^-,\Lambda,C)$, out of which we construct the components of
$\mathbb{T}(\bar \Phi_{w=0})$, denoted by
$(A,\Psi,B,G^-,\Lambda,C)\vert_{\mathbb{T}(\bar\Phi)}$.
In \cite{deWit:2010za} the following relation was established,
\begin{align}
  \label{eq:T-components}
  A\vert_{\mathbb{T}(\bar\Phi)} =&\, \bar C \,, \nonumber\\[.3ex]
  \Psi_i\vert_{\mathbb{T}(\bar\Phi)}=&\,-2\,\varepsilon_{ij}
  \Slash{D}\Lambda^j
  -6\, \;\varepsilon_{ik} \varepsilon_{jl} \chi^j B^{kl}
  -\tfrac14\varepsilon_{ij}\varepsilon_{kl} \, \gamma^{ab} T_{ab}{}^{jk}
  \stackrel{\leftrightarrow}{\Slash{D}}\Psi^l \,,  \nonumber\\[.6ex]
  B_{ij}\vert_{\mathbb{T}(\bar\Phi)} =&\, - 2\,\varepsilon_{ik}\varepsilon_{jl}
  \big(\Box_\mathrm{c} + 3\,D\big) B^{kl}   -2\, G^+_{ab}\,
  R(\mathcal{V})^{ab\,k}{}_{i}\, \varepsilon_{jk} 
  \,, \nonumber \displaybreak[0]\\[.6ex]
  G_{ab}^-\vert_{\mathbb{T}(\bar\Phi)} =&\, - \big(\delta_a{}^{[c} \delta_b{}^{d]}
    -\tfrac12\varepsilon_{ab}{}^{cd}\big)
    \big[4\, D_cD^e G^+_{ed} + (D^e\bar A
    \,D_cT_{de}{}^{ij}+D_c\bar A
   \,D^eT_{ed}{}^{ij})\varepsilon  _{ij} \big] \nonumber\\
  &\, +\Box_\mathrm{c} \bar A \,T_{ab}{}^{ij}\varepsilon_{ij}
    -R(\mathcal{V})^-{}_{\!\!ab}{}^i{}_k \,B^{jk} \,\varepsilon_{ij}
    +\tfrac1{8} T_{ab}{}^{ij} \,T_{cdij} G^{+cd} 
    \,, \nonumber \displaybreak[0]\\[.6ex]
    \Lambda_i\vert_{\mathbb{T}(\bar\Phi)} =&\, 2\,\Box_\mathrm{c}\Slash{D}
    \Psi^{j}\varepsilon_{ij}
    + \tfrac1{4}  \gamma^c\gamma_{ab} (2\, D_c
      T^{ab}{}_{ij}\,\Lambda^{j} + T^{ab}{}_{ij} \,D_c \Lambda^{j})
      \nonumber\\
   &\,
   - \tfrac1{2}\varepsilon_{ij} \big(R(\mathcal{V})_{ab}{}^j{}_k +
   2\mathrm{i} \, R(A)_{ab}\delta^j{}_k\big)\,\gamma^c\gamma^{ab} D_c\Psi^k
   \nonumber\\
   &\,
   +\tfrac12\,\varepsilon_{ij} \big( 3\, D_b D
   - 4\mathrm{i} D^a R(A)_{ab}
     +\tfrac{1}{4}
     T_{bc}{}^{ij}\stackrel{\leftrightarrow}{D_a} T^{ac}{}_{ij}
     \big)\,\gamma^b \Psi^{j}  \nonumber\\
   &\,
   -2\,G^{+ab}\, \Slash{D}R(Q)_{ab}{}_{i}
   +6\,\varepsilon_{ij}  D\, \Slash{D} \Psi^{j}  \nonumber\\
   &\,       + 3 \,\varepsilon_{ij}\,\big(\Slash{D}\chi_k\,B^{kj}
     +\Slash{D} \bar A\,\Slash{D}\chi^{j}  \big)  \nonumber\\
   &\,
   + \tfrac32\big( 2\,\Slash{D}B^{kj} \varepsilon_{ij}
    +\,  \Slash{D} G_{ab}^+ \gamma^{ab} \, \delta^k_{i}
    +\tfrac14  \varepsilon_{mn} T_{ab}{}^{mn}\,\gamma^{ab}
    \,\Slash{D}\bar A\, \delta_i{}^k\big)  \chi_k 
    \,,  \nonumber \displaybreak[0]\\[.6ex]
  C\vert_{\mathbb{T}(\bar\Phi)}=&\,
  4(\Box_\mathrm{c} + 3\, D) \Box_\mathrm{c} \bar A -\tfrac12
  D_a\big(T^{ab}{}_{ij} \,T_{cb}{}^{ij}\big) \,D^c\bar A
   +\tfrac1{16} (T_{abij}\varepsilon^{ij})^2 \bar C
  \nonumber\\
  &\,
  + D_a\big(\varepsilon^{ij} D^a T_{bc ij}\,G^{+bc} +4
  \,\varepsilon^{ij} T^{ab}{}_{ij} \,D^cG^{+}_{cb} -
     T_{bc}{}^{ij}\, T^{ac}{}_{ij} \,D^b\bar A \big)  \nonumber\\
  &\,
  + \big( 6\,D_b D   - 8\mathrm{i}D^a R(A)_{ab} \big) \,D^b \bar A
  + \,,
\end{align}
where we suppressed terms nonlinear in the
covariant fermion fields. Observe that the right-hand side of
these expressions is always linear in the conjugate components of the
$w=0$ chiral multiplet, i.e. in $(\bar A,\Psi^i,B^{ij}, G^+_{ab},
\Lambda^i, \bar C)$.

Using the result \eqref{eq:T-components} one can construct a large
variety of superconformal invariants with higher-derivative couplings
involving vector multiplets, as well as the tensor and Weyl chiral
backgrounds. The construction of
the higher-order Lagrangians therefore proceeds in two steps. First one
constructs the Lagrangian in terms of unrestricted chiral multiplets of
appropriate Weyl weights, in the form
\begin{equation}
  \label{eq:general-kin}
  \int \;\mathrm{d}^4\theta \,\Phi_0 \,\mathbb{T}^{(n_1)}\,\mathbb{T}^{(n_2)}
  \cdots \,\mathbb{T}^{(n_k)} \,.
\end{equation}
Here, the $n$-th power of the kinetic multiplet is defined recursively as
$\mathbb{T}^{(n)}= \mathbb{T}(\bar\Phi_n \,\mathbb{T}^{(n-1)})$ for
$\Phi_n$ of appropriate weight. Subsequently, one expresses the
unrestricted supermultiplets in terms of the reduced supermultiplets in
section \ref{App:4D-chiral-multiplets}. In these expressions it is natural to
introduce a variety of arbitrary homogeneous functions, so that resulting
final Lagrangian is controlled by a function of given homogeneity and
holomorphicity in the various fields, corresponding to the original structure
in \eqref{eq:general-kin}.

In this work, we will make use of invariants of the type \eqref{eq:general-kin},
where one, two or three kinetic multiplets appear, and are naturally quadratic,
cubic and quartic in chiral multiplet components, respectively. While the first
of these was described in detail in \cite{deWit:2010za}, the other two have not
appeared in the literature. These are straightforward to write, using the formulae
above and in \cite{deWit:2010za} but are rather unilluminating,
so that we prefer to emphasise the structure of the corresponding Lagrangians,
restricting ourselves to the leading terms.

\subsection*{The quadratic invariant}

The simplest case of a Lagrangian involving a kinetic multiplet is the one
in \eqref{eq:real-chiral-n2}, where a $w=0$ chiral multiplet is multiplied
with the kinetic of an antichiral one. In components, the leading bosonic
terms in the resulting Lagrangian read
\begin{align}
  \label{eq:quadratic-chiral-Lagr}
  e^{-1}\mathcal{L} =&\,
  C\,\bar C + 8\, \mathcal{D}_a F^{-ab}\, \mathcal{D}^c F^+{}_{cb}
   + 4\, F^{-ac}\, F^+{}_{bc}\, R(\omega,e)_a{}^b
     \nonumber \\[.1ex]
  &\,
  +4\,\mathcal{D}^2 A\,\mathcal{D}^2\bar A
  + 8\,\mathcal{D}^\mu A\, \big[R_\mu{}^a(\omega,e) -\tfrac13
  R(\omega,e)\,e_\mu{}^a \big]\mathcal{D}_a\bar A
  \nonumber \\[.1ex]
  &\,
   - \mathcal{D}^\mu B_{ij} \,\mathcal{D}_\mu B^{ij} + (\tfrac16
   R(\omega,e) +2\,D) \,
   B_{ij} B^{ij}
 + \cdots
\,,
\end{align}
where we suppressed the prime on the second chiral multiplet
indicated in \eqref{eq:real-chiral-n2} for brevity.
The next step is to consider the components of the chiral and anti-chiral
multiplet in \eqref{eq:quadratic-chiral-Lagr} to be composite, given as
holomorphic and anti-holomorphic functions, $F$, $\bar F$ of the fundamental
vector, tensor and Weyl multiplet respectively. The result is a Lagrangian that
is controlled by a homogeneous function of degree zero,
\begin{equation}\label{eq:patch-functions}
 F(X^A, A_{\sf w}, A_{\sf t})\, \bar{F}(\bar{X}^A, \bar{A}_{\sf w}, \bar{A}_{\sf t})
 \sim \cH( X^A, A_{\sf w}, A_{\sf t}, \bar{X}^A, \bar{A}_{\sf w}, \bar{A}_{\sf t} )\,,
\end{equation}
which depends on the vector multiplets scalars, $X^A$, and the
Weyl and tensor multiplet composites, $A_{\sf w}$ and $A_{\sf t}$.
This invariant corresponds to higher derivative couplings that are
quadratic in the leading terms, $F^2$, $R^2$ and $(\nabla E)^2$ respectively.
The arbitrariness of the function in $A_{\sf w}$ is analogous to the similar
dependence of the chiral couplings, $F(X^A, A_{\sf w})$ which describes the
full topological string partition function. 
Note that the various combinations have different order of derivatives,
as e.g. $F^4$ comprises only four derivatives, while $R^2 F^2$,
$(\nabla E)^2 F^2$ contain six derivatives and $R^4$, $R^2 (\nabla E)^2$,
$(\nabla E)^4$ contain eight derivatives. However, all these invariants
have a common structure, found by substituting the definitions of the
chiral multiplets in terms of $F$, $\bar F$ and $\cH$ in
\eqref{eq:quadratic-chiral-Lagr}.

This was done in \cite{deWit:2010za}, where the $F^4$ coupling was
constructed, based on a real function $\cH(X,\bar X)$, which plays the
role of a K\"ahler potential, as it is defined up to a real function, as
\begin{equation}
  \label{eq:kahler}
  \mathcal{H}(X,\bar X)\to \mathcal{H}(X,\bar X) +
  \Lambda(X)+\bar\Lambda(\bar X)\,.
\end{equation}
The explicit form of the Lagrangian is
\begin{align}
  \label{eq:real-susp-action}
  e^{-1}\mathcal{L} =&\, \mathcal{H}_{IJ\bar K \bar L}\Big[\tfrac14
    \big( G_{ab}^-{}^I\, G^{-ab\,J}
                -\tfrac12 Y_{ij}{}^I\, Y^{ijJ} \big)
                \big( G_{ab}^+{}^K \, G^{+ab\,L} -\tfrac12 Y^{ijK}\,
                  Y_{ij}{}^L  \big)
              \nonumber\\
              & \qquad\quad +4\,\mathcal{D}_a X^I\, \mathcal{D}_b \bar X^K
                \big(\mathcal{D}^a X^J \,\mathcal{D}^b \bar X^L
                  + 2\, G^{-\,ac\,J}\,G^{+\,b}{}_c{}^L -
                  \tfrac14 \eta^{ab}\, Y^J_{ij}\,Y^{L\,ij}\big)
              \Big]\nonumber\\[.5ex]
   +&\,\Big\{ \mathcal{H}_{IJ\bar K}\Big[4\,\mathcal{D}_a X^I\,
     \mathcal{D}^a X^J\, \mathcal{D}^2\bar X^K
       - \mathcal{D}_a
            X^I\, Y^J_{ij}\,\mathcal{D}^aY^{K\,ij} \nonumber\\
  & \qquad\quad - \big(G^{-ab\,I}\, G_{ab}^{-\,J} -\tfrac12 Y^I_{ij}\,
Y^{Jij})
      \big( \Box_\mathrm{c} X^K + \tfrac18 G^{-\,K}_{ab}\, T^{ab ij}
          \varepsilon_{ij}\big) \nonumber\\[.5ex]
  & \qquad\quad +8 \,\mathcal{D}^a X^I G^{-\,J}_{ab}
  \big( \mathcal{D}_cG^{+\,cb\,K}- \tfrac12 \mathcal{D}_c\bar X^K
              T^{ij\,cb} \varepsilon_{ij}\big) \Big]
            +\mathrm{h.c.}\Big\}   \displaybreak[0] \nonumber\\[.5ex]
     +&\mathcal{H}_{I\bar J}\Big[ 4\big( \Box_\mathrm{c} \bar X^I + \tfrac18
         G_{ab}^{+\,I}\, T^{ab}{}_{ij} \varepsilon^{ij}\big)
     \big( \Box_\mathrm{c}  X^J + \tfrac18 G_{ab}^{-\,J}\, T^{abij}
       \varepsilon_{ij}\big) + 4\,\mathcal{D}^2 X^I \,\mathcal{D}^2
       \bar X^J \nonumber\\
       & \quad\quad +8\,\mathcal{D}_{a}G^{-\,abI\,}\,
       \mathcal{D}_cG^{+c}{}_{b}{}^J   - \mathcal{D}_a Y_{ij}{}^I\,
            \mathcal{D}^a Y^{ij\,J}
            +\tfrac1{4} T_{ab}{}^{ij} \,T_{cdij}
            \,G^{-ab\,I}G^{+cd\,J}
     \nonumber\\
     &\quad\quad
     +\big(\tfrac16 \mathcal{R} +2\,D\big) Y_{ij}{}^I\, Y^{ij\,J}   + 4\,
     G^{-ac\,I}\, G^{+}{}_{bc}{}^J \, \mathcal{R}_a{}^b  \nonumber\\
     &\quad\quad + 8\big(\mathcal{R}^{\mu\nu}-\tfrac13 g^{\mu\nu}
     \mathcal{R} +\tfrac1{4} T^\mu{}_{b}{}^{ij}\, T^{\nu b}{}_{ij}
     +\mathrm{i} R(A)^{\mu\nu} - g^{\mu\nu} D\big) \mathcal{D}_\mu X^I
     \,\mathcal{D}_\nu \bar X^J  \nonumber\\
     &\quad\quad
     - \big[\mathcal{D}_c \bar X^J \big(\mathcal{D}^c
     T_{ab}{}^{ij}\,G^{-\,I\,ab} +4
       \,T^{ij\,cb} \,\mathcal{D}^aG^{-\,I}_{ab} \big)\varepsilon_{ij}
       +[\mathrm{h.c.}; I\leftrightarrow J]  \big]\nonumber\\
     &\quad\quad -\big[\varepsilon^{ik}\, Y_{ij}{}^I\, G^{+ab\,J}\,
      R(\mathcal{V})_{ab}{}^j{}_k +[\mathrm{h.c.}; I\leftrightarrow J]
      \big]  \Big] \,,
\end{align}
where (we suppress fermionic contributions),
\begin{align}
  \label{eq:def}
  G_{ab}^-{}^I =&\,  F^-_{ab}{}^I
  -\tfrac14\, \bar{X}^I\, T_{ab}{}^{ij}\varepsilon_{ij} \,,
  \nonumber\\
  \Box_\mathrm{c} X^I=&\, \mathcal{D}^2 X^I + \big(\tfrac16
  \mathcal{R} +D\big) \,X^I \,.
\end{align}

One can obtain the more general couplings as discussed above, resulting in similar
expressions. For example, the $R^2F^2$- and $R^4$-type couplings feature terms
found by substituting $F^2\rightarrow R^2$ and similarly for the other components
in \eqref{eq:real-susp-action} and are discussed in \cite{deWit:2010za}. 

\subsection*{The cubic invariant}

The next more complicated example of Lagrangians containing kinetic multiplets
is to consider an integral quadratic in kinetic multiplets, as
\begin{equation}
  \label{eq:chiral-n3}
  \int \;\mathrm{d}^4\bar\theta \,\bar\Phi_0\mathbb{T}(\Phi_1)\,
    \mathbb{T}( \Phi_2) \,,
\end{equation}
where $\Phi_0$ is a $w=-2$ chiral, while $\Phi_1$ and $\Phi_2$ are $w=0$
anti-chirals, as above. It is straightforward to apply the multiplication rule
for chiral multiplets, to obtain the analogous master formula of the type
\eqref{eq:quadratic-chiral-Lagr}, in this case. The result takes the form
\begin{align}
  \label{eq:cubic-chiral-Lagr}
  e^{-1}\mathcal{L} =&\,
  C^2\,\bar C - \tfrac1{16}\,\bar A\,C^2 (T^+)^2
     \nonumber \\[.1ex]
  &\,
  + 2\,\bar{A}\, C\,\big[
   4\,(\Box + 3\,D)\Box A + \tfrac1{16}\,C\, (T^-)^2
  +4\, D_a\big( T^{+\,ab} \,D^cG^{+}_{cb}\big)
   + \dots \big]
     \nonumber \\[.1ex]
  &\,
  - \bar{A}\, C\,\big[
   2\,\varepsilon^{ik}\varepsilon^{jl}(\Box + 3\,D)B_{ij}\,(\Box + 3\,D)B_{kl}
   - D_{[a}\big(D^c G^{-}_{c b]_+} \big) D^{[a}\big(D_c G^{-}{}^{c b]_+} \big)
   + \dots \big]
     \nonumber \\[.1ex]
  &\,
  + 2\,C\,\big[ B^{ij}\,(\Box + 3\,D)B_{ij}
  - G^{+\,ab} \left( D_{[a}\big(D^c G^{-}_{c b]_+} \big) - \Box A\,T^+_{ab} \right)
  + \cdots \big]
\,,
\end{align}
which is manifestly quadratic in holomorphic and linear in anti-holomorphic components.
Note that we again use a simplified notation that naively identifies the three a priori
independent multiplets, despite the fact that the anti-chiral multiplet is of weight
$-2$, while the chiral ones are of $w=0$. The most general invariant follows by completing
the combinations given above with the components of the kinetic multiplet given in
\eqref{eq:T-components} and viewing the holomorphic components as quadratic forms in
the components of the two chiral multiplets in \eqref{eq:chiral-n3}, as done in
\eqref{eq:quadratic-chiral-Lagr}.

It is now straightforward, if cumbersome, to consider the three multiplets in
\eqref{eq:chiral-n3} as functions of the vector multiplets, the tensor multiplet
and the Weyl multiplet, as done in \eqref{eq:patch-functions}, leading to a Lagrangian
described by a function, $\cH( X^A, A_{\sf w}, A_{\sf t}, \bar{X}^A, \bar{A}_{\sf w}, \bar{A}_{\sf t} )$,
which is homogeneous of degree zero in the holomorphic components and homogeneous of
degree $-2$ in the anti-holomorphic components. We refrain from giving the corresponding
expression \eqref{eq:real-susp-action} in this case, since we will only be dealing with
the leading terms and the properties of the corresponding function $\cH$.

Once again, the generic function of all available multiplets leads to various
invariants, which contain different orders of derivatives but share the same structure,
as in \eqref{eq:cubic-chiral-Lagr}. The prototype of these terms is the $F^6$ invariant
arising by taking $\cH( X^A, \bar{X}^A)$, i.e. a function of vector multiplet scalars
only. Allowing for holomorphic/anti-holomorphic dependence on the scalars $A_{\sf w}$
and $A_{\sf t}$ leads to terms of the type $R^2 F^4$, $R^4 F^2$, $(\nabla E)^2 F^4$ and
so on for all possible combinations. Note that many of these contain more than eight
derivatives and therefore fall outside the scope of this work.

\subsection*{The quartic invariants}

We finally consider integrals of the type \eqref{eq:general-kin} which are cubic in
the kinetic multiplet operator, $ \mathbb{T}$, in which case we find two possibilities.
Indeed, this is the first case where one needs to consider nested kinetic multiplets,
since the two possible integrals,
\begin{equation}
  \label{eq:chiral-n4}
  \int \;\mathrm{d}^4\bar\theta \,\bar\Phi_0\mathbb{T}(\Phi_1)\,
    \mathbb{T}(\Phi_2)\, \mathbb{T}(\Phi_3) \,,
\qquad
  \int \;\mathrm{d}^4\bar\theta \,\bar\Phi_0\mathbb{T}(\Phi_1)\,
    \mathbb{T}( \Phi^\prime_0 \mathbb{T}(\bar \Phi_2) )\,,
\end{equation}
are not equivalent upon partial integration. Here, the first integral is the
straightforward extension of \eqref{eq:real-chiral-n2} and \eqref{eq:chiral-n3}, while in the
second integral $\Phi_0$ and $\Phi^\prime_0$ are $w=-2$ chirals, while $\Phi_1$ and
$\Phi_2$ are $w=0$ chirals, as above.

Once again, one can apply the multiplication rule for chiral multiplets, to obtain
the analogous master formula of the type \eqref{eq:quadratic-chiral-Lagr}, in these
cases. The expression for the first integral is similar to \eqref{eq:cubic-chiral-Lagr},
where three chiral multiplets appear and is not used in this paper.
The second integral is more cumbersome, but can be easily computed by an iterative
procedure, by noting that $\bar\Phi_0\mathbb{T}(\Phi_1)$ and
$\Phi^\prime_0 \mathbb{T}(\bar \Phi_2)$ are $w=0$ multiplets, so that
\eqref{eq:quadratic-chiral-Lagr} applies for their components. One can then
obtain the result to the integral by making the following substitutions
\begin{align}
 A \rightarrow &\, A_0\, A\vert_{\mathbb{T}(\bar\Phi)}\,,
\nonumber\\
 B_{ij} \rightarrow &\, B_{0\,ij}\, A\vert_{\mathbb{T}(\bar\Phi)} + A_0\,B_{ij}\vert_{\mathbb{T}(\bar\Phi)}\,,
\nonumber\\
 G^{-\,ab} \rightarrow &\, G^{-\,ab}_{0}\, A\vert_{\mathbb{T}(\bar\Phi)} + A_0\,G^{-\,ab}\vert_{\mathbb{T}(\bar\Phi)}\,,
\nonumber\\
 C \rightarrow &\, C_{0}\, A\vert_{\mathbb{T}(\bar\Phi)} + A_0\,C\vert_{\mathbb{T}(\bar\Phi)}
  - \tfrac14\,\big( \varepsilon^{ik}\varepsilon^{jl}B_{0\,ij} \,B_{kl}\vert_{\mathbb{T}(\bar\Phi)}
  - 2\, G^-_{0\,ab} \,G^{-\,ab}\vert_{\mathbb{T}(\bar\Phi)} \big)\,,
\end{align}
in \eqref{eq:quadratic-chiral-Lagr}, where the components labeled with $\vert_{\mathbb{T}(\bar\Phi)}$
are as in \eqref{eq:T-components}.

As above, allowing for the four chiral multiplets involved
to depend on the vector, tensor and/or the Weyl multiplet, exactly as in \eqref{eq:patch-functions},
one obtains various higher derivative invariants, sharing the same structure. However, all but one of
the invariants described by each of the two integrals in \eqref{eq:chiral-n4} necessarily contain more than eight
spacetime derivatives if the Weyl and tensor multiplet backgrounds are allowed, so that they are not relevant
for our consideration. The exception is the case where all the composite chiral multiplets only depend on the vector multiplets, in which case we obtain two $F^8$ invariants from \eqref{eq:chiral-n4}.

\end{appendix}

\bibliographystyle{utphys}
\bibliography{CYReduction}

\end{document}